\documentclass[12pt]{article}
\pdfoutput=1

\usepackage{amsmath,amssymb,amscd}
\usepackage{listings}
\usepackage{caption}
\usepackage{dsfont}
\usepackage{slashed}
\usepackage{color}

\usepackage[pdftex]{graphicx}
\usepackage{epstopdf}
\usepackage{subfigure}
\usepackage{epsfig}
\usepackage{listings}
\usepackage{caption}
\usepackage{cite}

\usepackage{multirow}

\newcommand{\bea}{\begin{align}}
\newcommand{\eea}{\end{align}}
\newcommand{\beq}{\begin{equation}}
\newcommand{\eeq}{\end{equation}}
\newcommand{\nbea}{\begin{align*}}
\newcommand{\neea}{\end{align*}}
\newcommand{\nbeq}{\begin{equation*}}
\newcommand{\neeq}{\end{equation*}}
\newcommand{\bear}{\begin{eqnarray}}  
\newcommand{\eear}{\end{eqnarray}}  

 \def\lra#1{\overset{\text{\scriptsize$\leftrightarrow$}}{#1}}

\numberwithin{equation}{section}
\begin{document}

%\begin{titlepage}

%\pagestyle{empty}

\baselineskip=21pt
\rightline{KCL-PH-TH/2014-41, LCTS/2014-41, CERN-PH-TH/2014-201}
\vskip 1in

\begin{center}

{\large {\bf The Effective Standard Model after LHC Run I }}

\vskip 0.6in

 {\bf John~Ellis}$^{1,2}$,~%\footnote{John.Ellis@cern.ch},
 {\bf Ver\'onica Sanz}$^{3}$~%\footnote{v.sanz@sussex.ac.uk}
and {\bf Tevong~You}$^{1}$%\footnote{tevong.you@kcl.ac.uk}

\vskip 0.4in

{\small {\it

$^1${Theoretical Particle Physics and Cosmology Group, Physics Department, \\
King's College London, London WC2R 2LS, UK}\\
$^2${TH Division, Physics Department, CERN, CH-1211 Geneva 23, Switzerland}\\
$^3${Department of Physics and Astronomy, University of Sussex, Brighton BN1 9QH, UK}\\
}}

\vskip 0.75in

{\bf Abstract}

\end{center}

\baselineskip=18pt \noindent

%%%%%%%%%%%%%%%%%%%%%%%%%%%%%%%%%%%%%%%%%%%%%%%%%

{\small
%JE
We treat the Standard Model as the low-energy limit of an effective field theory that incorporates higher-dimensional
operators to capture the effects of decoupled new physics. We consider the constraints
imposed on the coefficients of dimension-6 operators by electroweak precision tests (EWPTs),
applying a framework for the effects of dimension-6 operators
on electroweak precision tests that is more general than the standard $S,T$ formalism, and use
measurements of Higgs couplings and the kinematics of associated Higgs production  at the Tevatron and LHC,
as well as triple-gauge couplings at the LHC.
We highlight the complementarity between EWPTs, Tevatron and LHC measurements in obtaining
model-independent limits on the effective Standard Model after LHC Run~1. We illustrate the
combined constraints with the example of the two-Higgs doublet model.

%%%%%%%%%%%%%%%%%%%%%%%%%%%%%%%%%%%%%%%%%%%%%%%%

\vskip 1in

\leftline{October 2014}

%\end{titlepage}
\newpage

%\tableofcontents

%*******************************************************************************************
\section{Introduction}

Run 1 of the LHC has taken probes of the Standard Model to a new level, not only by
the discovery of the Higgs boson $H(125)$~\cite{Eureka} and the absence of other new particles,
but also via the new constraints imposed on the couplings of vector bosons and the top quark~\cite{fits}.
Now is an appropriate time to assess the global constraints placed on possible new physics
by LHC Run 1 in conjunction with the Tevatron, LEP and other experiments. In view of the
kinematic reach of the LHC, it is natural to suppose that the threshold for any new physics
may lie substantially above the masses of the Standard Model particles. In this case, the
new physics may be analyzed in the decoupling limit~\cite{decoupling}, and its effects may be parameterized
in terms of higher-dimensional operators composed of Standard Model fields~\cite{eft}. Using the equations of motions reduces the number of independent operators~\cite{eomreduction}, with a complete non-redundant set first categorised in \cite{GIMR}. 

This is the effective Standard Model approach adopted in a large number of
recent papers\footnote{For earlier studies of dimension-6 operators in triple-gauge couplings and Higgs physics see for example \cite{earlyeft}.}~\cite{recenteftpapers, EFTreview}, and there have been many analyses of
the constraints imposed on new physics via upper limits on the coefficients of a complete dimension-6
operator basis~\cite{eftconstraints, HanSkiba, Ciuchinietal, PomarolRiva, ESY3}, in particular. Several different classes of measurements make important contributions
to these constraints. LEP and other experiments contribute via electroweak precision tests (EWPTs)~\cite{LEPEWPT},
which are often presented as constraints on the $S$ and $T$ parameters that are defined in
terms of oblique radiative corrections due to vacuum polarization diagrams, and via measurements
of triple-gauge couplings (TGCs). The Tevatron experiments contribute via measurements of
(constraints on) production of the Higgs boson $H$ in association with massive gauge bosons $V = W^\pm, Z^0$~\cite{D0comb}.
Finally, the LHC experiments contribute via many Higgs measurements including signal strengths~\cite{strengths},
branching ratios and kinematic distributions~\cite{atlasptv}, and also via TGC measurements~\cite{CMS78TGC, ATLAS8TGC}.

We demonstrated in previous work~\cite{ESY3} the power of the constraints provided by measurements
of kinematic distributions in $V + H$ production at the Tevatron and the LHC, showing that
measurements of the $V + H$ invariant mass $M_{VH}$ at the Tevatron and the transverse
momentum $p_T^V$ at the LHC could close off a `blind' direction in the parameter space of
dimension-6 operator coefficients that had been allowed by previous analyses of LEP and LHC data~\cite{leshouches}.
Subsequently, new data on TGCs from LHC running at 8 TeV have been published~\cite{CMS78TGC,ATLAS8TGC}.
In this paper we make the first complete analysis of the data from LHC Run~1 and the Tevatron,
in combination with the EWPT constraints, considering only CP-even operators and assuming minimal flavour violation. 
The dimension-6 operators we consider and the 95\% CL ranges that we find for their
coefficients are listed in Tables~\ref{tab:EWPToperators} and \ref{tab:LHCoperators}.

We confirm previous findings that the EWPTs place very strong constraints on certain
(combinations of) operator coefficients. On the other hand, we also find that the Higgs
observables (signal strengths and associated production kinematics) and the TGC
measurements at the LHC also have complementary r{\^ o}les to play. Some operator
coefficients are better constrained by the TGC data, and some by the Higgs data. One coefficient in particular only affects TGCs and nothing else. Only
their combination provides a complete picture of the constraints on the dimension-6
operator coefficients after LHC Run~1. 

The outline of this paper is as follows. In Section~2 we discuss the EWPTs, first
reviewing a general expansion formalism for EWPTs, and then demonstrating that it
reproduces the constraints on the vacuum polarization parameters $S$ and $T$ found in
other analyses before illustrating its use in capturing the effects of a complete basis of dimension-6 operators. In Section~3 we discuss the constraints imposed by measurements of Higgs
couplings, associated Higgs production kinematics and TGCs at the LHC, demonstrating
their complementarity.
Section~4 illustrates the application of these combined constraints on the
coefficients of dimension-6 operators to the two-Higgs-doublet model (2HDM).
Section~5 summarizes our conclusions and assesses some future prospects,
and an Appendix discusses aspects of kinematics and the applicability of
effective field theory in our analysis.

\section{Electroweak Precision Tests at LEP}

Electroweak precision tests (EWPTs), particularly those provided by LEP,
are amongst the most sensitive observables for constraining new physics beyond the Standard Model.
EWPTs are typically summarized via constraints on the $S$ and $T$ parameters~\cite{PeskinTakeuchi}
and their generalization to include the $W$ and $Y$ parameters~\cite{beyondSTU} that are relevant for 
custodially-symmetric and weak isospin-preserving new physics, which characterize the Standard Model
vector boson self-energy corrections~\footnote{See also~\cite{altarellibarbieri} for another parametrisation of 
EWPT fits that includes vertex corrections in a set of $\epsilon$ parameters.}. If new physics affects only the Standard Model
gauge sector and does not couple directly to Standard Model fermions, this approach
may be sufficient for placing bounds on such `universal' models, but the effective Standard
Model also includes fermionic operators that affect electroweak precision tests.
Thus a more general framework is required to capture all the possible effects of
decoupled new physics in a model-independent way. 

There have been many studies considering individual or subsets of bounds for all dimension-6 operators entering in EWPTs, for example \cite{barbieristrumia, 1303.3876continoetal}, and full analyses including simultaneously a complete basis of dimension-6 operators affecting these EWPTs have been
performed in~\cite{HanSkiba, Ciuchinietal, PomarolRiva}, but a full calculation of the effects of propagation
of corrections to input observables and self-energies as well as direct contributions to observables
was needed in each different basis. Here we employ instead the recent expansion formalism of~\cite{wellsandzhang},
which separates the calculation of the corrections' effects on the EWPT observables and the
calculations of the contributions to the corrections from new physics. This framework facilitates any
$\chi^2$ analysis that seeks to go beyond the $S,T$ parametrization and renders
more transparent the origin of the effects from each operator.

\subsection{The Expansion Formalism} 

For convenience, we briefly summarize here the analysis of~\cite{wellsandzhang}.
The principle is that, given the Standard Model with Lagrangian parameters
$p_\text{SM} \equiv \{ g, g^\prime, g_s, y_t, v, \lambda \}$, one may calculate theoretical values
$\hat{\mathcal{O}}^\text{th}_i(p_\text{SM})$ for the observables
\begin{equation*}
\hat{\mathcal{O}}_i \equiv \{ m_Z, G_F, \alpha(m_Z), m_t, \alpha_s, m_H, m_W, \Gamma_l, \Gamma_q, \sigma_\text{had}, R_l, \sin^2{\theta_\text{eff}}, A_f, A^f_{FB}, ...\}
\end{equation*}
that are measured by experiments with errors $\Delta\hat{\mathcal{O}}^\text{exp}_i$. To compare the theoretical predictions
$\hat{\mathcal{O}}^\text{th}_i(p_\text{SM})$ with the experimental measurements, $\hat{\mathcal{O}}^\text{exp}_i$, 
we must first choose 6 of these observables as `input' observables $\hat{\mathcal{O}}_{i^\prime}$, 
typically the most precisely measured ones~\footnote{Another convenient choice of input observables is to use $m_W$ instead of $G_F$~\cite{BSMprimaries}.}, such as 
\begin{equation*}
\hat{\mathcal{O}}_{i^\prime} \equiv \{ m_Z, G_F, \alpha(m_Z), m_t, \alpha_s, m_H \}	\, .
\end{equation*}
These assign values $p_\text{SM}^\text{ref}$ to the Lagrangian parameters such that the
$\hat{\mathcal{O}}^\text{th}_{i^\prime}(p_\text{SM}^\text{ref})$ agree well with measurements,
and numerical values for the other `output' observables can then be obtained in terms of $p^\text{ref}_\text{SM}$. 

In the presence of new physics characterized by parameters $p_\alpha$, 
the theoretical expressions for the observables are modified by a correction
$\delta^\text{NP}\hat{\mathcal{O}}_i(p_\text{SM},p_\alpha)$:
\begin{equation*}
\hat{\mathcal{O}}_i^\text{th}(p_\text{SM},p_\alpha) = \hat{\mathcal{O}}_i^\text{SM}(p_\text{SM}) + \delta^\text{NP}\hat{\mathcal{O}}_i(p_\text{SM},p_\alpha) 	\, .
\end{equation*}
Since the relations between input observables and Lagrangian parameters are modified in general,
a different $p_\text{SM}^\text{ref}$ value would normally be preferred to compensate for
non-zero values of $p_\alpha$ so as to remain in agreement with experiment.
This may be quantified by a $\chi^2$ analysis that varies the parameters
$(p_\text{SM},p_\alpha)$ so as to minimize the function
\begin{equation*}
\chi^2(p_\text{SM},p_\alpha) = \sum_{i,j} (\hat{\mathcal{O}}_i^\text{th} - \hat{\mathcal{O}}_i^\text{exp})(\sigma^2)^{-1}_{ij}(\hat{\mathcal{O}}_j^\text{th} - \hat{\mathcal{O}}_j^\text{exp})	\quad , \quad (\sigma^2)_{ij} = \Delta\hat{\mathcal{O}}_i^\text{exp}\rho_{ij}\Delta\hat{\mathcal{O}}_j^\text{exp}	\, ,
\end{equation*}
where $\rho_{ij}$ is the correlation matrix. 

To avoid recomputing the full expression $\hat{\mathcal{O}}_i^\text{th}(p_\text{SM},p_\alpha)$ for each value of 
$p_\text{SM}$ and $p_\alpha$, the expansion formalism involves expanding about the Standard Model
reference values for the Lagrangian parameters:
\begin{align*}
\hat{\mathcal{O}}_i^\text{SM}(p_\text{SM}) &= \hat{\mathcal{O}}^\text{SM}_i(p_\text{SM}^\text{ref}) + \sum_{p_\text{SM}}
\frac{\partial\hat{\mathcal{O}}_i^\text{SM}}{\partial p_\text{SM}} (p_\text{SM} - p_\text{SM}^\text{ref}) + \text{...}	\\
&\simeq \hat{\mathcal{O}}^\text{ref}_i [1 + \bar{\delta}^\text{SM}\hat{\mathcal{O}}_i(p_\text{SM})]	\, ,
\end{align*}
where $\hat{\mathcal{O}}_i^\text{ref} \equiv  \hat{\mathcal{O}}^\text{SM}_i(p_\text{SM}^\text{ref})$, 
$\bar{\delta}^\text{SM}\hat{\mathcal{O}}_i(p_\text{SM}) = \sum_{p_\text{SM}} G_{ip_\text{SM}} \bar{\delta}p_\text{SM}$, 
and the quantities $G_{ik^\prime} \equiv \frac{p_\text{SM}^\text{ref}}{\hat{\mathcal{O}}^\text{ref}_i}
\frac{\partial\hat{\mathcal{O}}_i^\text{SM}}{\partial p_\text{SM}}$ are expansion coefficients 
that need only to be calculated once. Here $\bar{\delta}p_\text{SM} \equiv 
(p_\text{SM} - p_\text{SM}^\text{ref})/p_\text{SM}^\text{ref}$, and the fractional shift $\bar{\delta}$ is defined in general as
$\bar{\delta}\hat{\mathcal{O}}_i \equiv (\hat{\mathcal{O}}_i - \hat{\mathcal{O}}_i^\text{ref})/\hat{\mathcal{O}}_i^\text{ref}$.
Furthermore, to emphasize that the $p_\text{SM}$ are not directly measurable,
but are determined from the input observables $\hat{\mathcal{O}}_{i^\prime}$,
we note that the Lagrangian parameters can be eliminated in favour of the input observables by inverting the relation
$\bar{\delta}^\text{SM}\hat{\mathcal{O}}_{i^\prime} =  \sum_{p_\text{SM}} G_{i^\prime p_\text{SM}}\bar{\delta}p_\text{SM}$,
so that
\begin{equation}
\bar{\delta}^\text{SM}\hat{\mathcal{O}}_i =  \sum_{i^\prime} G_{ip_\text{SM}} \left( \sum_{p_\text{SM}} (G^{-1})_{p_\text{SM} i^\prime} \bar{\delta}^\text{SM}\hat{\mathcal{O}}_{i^\prime}\right) = \sum_{i^\prime} d_{i i^\prime} \bar{\delta}^\text{SM}\hat{\mathcal{O}}_{i^\prime}	\, .
\label{eq:selfenergiesST}
\end{equation}
The expansion coefficients for the output observables in terms of input observables are then given by the matrix 
$d_{i i^\prime} \equiv \sum_{p_\text{SM}} G_{i p_\text{SM}} (G^{-1})_{p_\text{SM} i^\prime} $. 

The theoretical predictions for the output observables can now be written as 
$\hat{\mathcal{O}}_i^\text{th} = \hat{\mathcal{O}}_i^\text{ref}(1 + \bar{\delta}\hat{\mathcal{O}}_i^\text{th})$, with 
\begin{equation*}
\bar{\delta}\hat{\mathcal{O}}_i^\text{th} = \sum_{i^\prime} d_{i i^\prime} \bar{\delta}^\text{SM} \hat{\mathcal{O}}_{i^\prime} + \xi_i  =  \sum_{i^\prime} d_{i i^\prime}(\bar{\delta}\hat{\mathcal{O}}_{i^\prime}^\text{th} - \xi_{i^\prime}) + \xi_i 	\, ,
\end{equation*}
where we used $\bar{\delta}\hat{\mathcal{O}}_{i^\prime}^\text{SM} = \bar{\delta}\hat{\mathcal{O}}_{i^\prime}^\text{th} - \xi_{i^\prime}$ and defined $\xi_i \equiv \delta^\text{NP}\hat{\mathcal{O}}_i / \hat{\mathcal{O}}_i^\text{ref}$. The $d_{i i^\prime}$ matrix is pre-calculated and encapsulates the dependence of each output observable on each input observable,
so that one needs only to plug in the contribution due to new physics that affect the input observables, 
$\xi_{i^\prime}$, and those that directly affect the output observables, $\xi_i$.
We note that, for the case of vector boson self-energy corrections, the
$\pi_{VV} \equiv \{ \pi_{ZZ}, \pi^\prime_{ZZ}, \pi_{\gamma Z}, \pi^\prime_{\gamma\gamma}, \pi_{+-}, \pi^0_{WW} \}$
are defined as in~\cite{wellsandzhang}, and the contributions to output observables through $\xi_{i^\prime}$ and $\xi_i$ are summarized by the given $b_{i, VV}$ coefficients. We then have
\begin{equation*}
\bar{\delta}\hat{\mathcal{O}}_i^\text{th} = \sum_{i^\prime}d_{i i^\prime}\bar{\delta}\hat{\mathcal{O}}_{i^\prime}^\text{th} + \bar{\delta}^\text{NP}\hat{\mathcal{O}}_i \, ,
\end{equation*}
where
\begin{equation}
\bar{\delta}^\text{NP}\hat{\mathcal{O}}_i \equiv \xi_i - \sum_{i^\prime} d_{i i^\prime}\xi_{i^\prime} + \sum_{VV} b_{i,VV}\delta^\text{NP}\pi_{VV}	\, ,
\label{eq:obsshift}
\end{equation}
and it remains only to determine the $\xi_{i^\prime}, \xi_i$ and $\delta^\text{NP}\pi_{VV}$ from
the dimension-6 operators in the effective Standard Model.

\subsection{Dimension-6 Operators in EWPTs}

\begin{table}[h]
\begin{center}
\begin{tabular}
{|c | c | c | c | c | c |}
\hline
\multirow{2}{*}{Operator} & \multirow{2}{*}{Coefficient} & \multicolumn{2}{| c |}{LEP Constraints}  \\
 &  & Individual & Marginalized \\
\hline
${\mathcal O}_W=\frac{ig}{2}\left( H^\dagger  \sigma^a \lra {D^\mu} H \right )D^\nu  W_{\mu \nu}^a$ & \multirow{2}{*}{$\frac{m_W^2}{\Lambda^2}(c_W+c_B)$} & \multirow{2}{*}{$(-0.00055, 0.0005)$}  & \multirow{2}{*}{$(-0.0033,0.0018)$}   \\
${\mathcal O}_B=\frac{ig'}{2}\left( H^\dagger  \lra {D^\mu} H \right )\partial^\nu  B_{\mu \nu}$ &  &  &  \\
\hline
${\cal O}_T=\frac{1}{2}\left (H^\dagger {\lra{D}_\mu} H\right)^2$ & $\frac{v^2}{\Lambda^2}c_T$ & $(0,0.001)$  & $(-0.0043, 0.0033)$  \\
\hline
$\mathcal{O}_{LL}^{(3)\, l}=( \bar L_L \sigma^a\gamma^\mu L_L)\, (\bar L_L \sigma^a\gamma_\mu L_L)$ & $\frac{v^2}{\Lambda^2}c^{(3)l}_{LL}$ & $(0,0.001)$  & $(-0.0013,0.00075)$  \\
\hline
${\mathcal O}_R^e =
(i H^\dagger {\lra { D_\mu}} H)( \bar e_R\gamma^\mu e_R)$  & $\frac{v^2}{\Lambda^2}c^e_R$ & $(-0.0015,0.0005)$  & $(-0.0018,0.00025)$  \\
\hline
${\cal O}_{R}^u =
(i H^\dagger {\lra { D_\mu}} H)( \bar u_R\gamma^\mu u_R)$ & $\frac{v^2}{\Lambda^2}c^u_R$ & $(-0.0035,0.005)$  & $(-0.011,0.011)$  \\
\hline 
${\cal O}_{R}^d =
(i H^\dagger {\lra { D_\mu}} H)( \bar d_R\gamma^\mu d_R)$ & $\frac{v^2}{\Lambda^2}c^d_R$ & $(-0.0075,0.0035)$  & $(-0.042,0.0044)$ \\
\hline 
${\cal O}_{L}^{(3)\, q}=(i H^\dagger \sigma^a {\lra { D_\mu}} H)( \bar Q_L\sigma^a\gamma^\mu Q_L)$ & $\frac{v^2}{\Lambda^2}c^{(3)q}_L$ & $(-0.0005,0.001)$  & $(-0.0044,0.0044)$ \\
\hline 
${\cal O}_{L}^q=(i H^\dagger {\lra { D_\mu}} H)( \bar Q_L\gamma^\mu Q_L)$  & $\frac{v^2}{\Lambda^2}c^q_L$ & $(-0.0015,0.003)$  & $(-0.0019,0.0069)$  \\
\hline
\end{tabular}
\end{center}
\caption{\it List of operators and coefficients in our basis entering in EWPTs at LEP, together with $95\%$ CL bounds when individual coefficients are switched on one at a time, and marginalized in a simultaneous fit. For the first four coefficients we report the constraints from the leptonic observables, while the remaining coefficients also include the hadronic observables. }
\label{tab:EWPToperators}
\end{table}

We begin with the familiar $S,T$ parameters before generalizing to a complete dimension-6 operator basis.
The universal parts of new physics contributions are often parametrized as oblique corrections to vector boson
self-energies, which can be written in terms of gauge eigenstates as
\begin{equation*}
\mathcal{L}_\text{VV} = -{W^+}^\mu\pi_{+-}(p^2)W^-_\mu - \frac{1}{2}{W^3}^\mu\pi_{33}(p^2)W^3_\mu - {W^3}^\mu\pi_{3B}(p^2)B_\mu - \frac{1}{2}B^\mu\pi_{BB}(p^2)B_\mu 	\, ,
\end{equation*}
where $\pi_{VV}(p^2) = \pi_{VV}^\text{SM}(p^2) + \delta\pi_{VV}(p^2)$.
Making a Taylor expansion at the quadratic order to which dimension-6 operators can contribute:
\begin{equation*}
\pi_{VV}(p^2) = \pi_{VV}(0) + p^2\pi^\prime_{VV}(0) + \frac{1}{2}(p^2)^2\pi^{\prime\prime}_{VV}(0) + \text{...} 	\, ,
\end{equation*}
the usual $\hat{S}$ and $\hat{T}$ parameters~\footnote{These are related to the $S$ and $T$ parameters
defined in~\cite{PeskinTakeuchi} via $S = \frac{4\sin^2{\theta_W}}{\alpha(m_Z)}\hat{S} \approx 119 \hat{S}$ and $T =\frac{1}{\alpha(m_Z)}\hat{T} \approx 129 \hat{T}$.} can be defined as
\begin{equation*}
\hat{S} \equiv \frac{g}{g^\prime}\frac{\pi^\prime_{3B}(0)}{\pi^\prime_{+-}(0)}	\quad , \quad \hat{T} \equiv \frac{\pi_{+-}(0) -\pi_{33}(0)}{\pi_{+-}(0)}	\, .
\end{equation*}
Since $U(1)_Q$ symmetry is conserved, which requires $\pi_{\gamma\gamma}(0)$ and $\pi_{\gamma Z}(0)$ to vanish by gauge invariance, the following relations must hold:
\begin{align*}
{g^\prime}^2\pi_{33}(0) + g^2\pi_{BB}(0) + 2gg^\prime \pi_{3B}(0) &= 0	\\
g\pi_{BB}(0) + g^\prime\pi_{3B}(0) &= 0 \, .
\end{align*}
After normalizing the $W^\pm$ and $B$ fields so that the kinetic terms are canonical and $\pi_{+-}(0) = -m_W^2$,
we obtain the following $\hat{S}$ and $\hat{T}$ corrections in the gauge mass eigenstates for the
quantities $\delta^\text{NP}\pi_{VV}$ defined in~\cite{wellsandzhang}:
\begin{align*}
\delta^\text{NP}\pi_{ZZ} &= -\hat{T} + 2\hat{S}\sin^2{\theta_W}	\\
\delta^\text{NP}\pi^\prime_{ZZ} &= 2\hat{S}\sin^2{\theta_W}	\\
\delta^\text{NP}\pi_{\gamma Z} &= -\hat{S}\cos{2\theta_W}\tan{\theta_W}	\\
\delta^\text{NP}\pi^\prime_{\gamma\gamma} &= -2\hat{S}\sin^2{\theta_W}	\, .
\end{align*}
Inserting these expressions into (\ref{eq:obsshift}) and performing a $\chi^2$ analysis in the expansion formalism,
using as output observables the EWPTs at the $Z$ peak and the $W$ mass: 
\begin{equation*}
\hat{\mathcal{O}}_i = \{ \Gamma_Z, \sigma^0_\text{had}, R^0_e, R^0_\mu, R^0_\tau, A^{0,e}_\text{FB}, \sin^2{\theta_\text{eff}^e}, R^0_b, R^0_c, A^{0,b}_\text{FB}, A^{0,c}_\text{FB}, A_b, A_c, \sin^2{\theta_\text{eff}^b}, \sin^2{\theta_\text{eff}^c}, m_W \}	\, ,
\end{equation*}
we obtain the $68\%, 95\%,$ and $99\%$ CL allowed regions for $S$ vs $T$ shown in Fig.~\ref{fig:ST},
denoted by dotted, dashed and solid contours respectively. We treat the observables as uncorrelated but
have checked that including the correlation matrix, for example in the leptonic subset as given in~\cite{LEPEWPT},
does not affect substantially our results, which agree reasonably closely with those of~\cite{1209.2716}.

\begin{figure}[h!]
\centering
\includegraphics[scale=0.5]{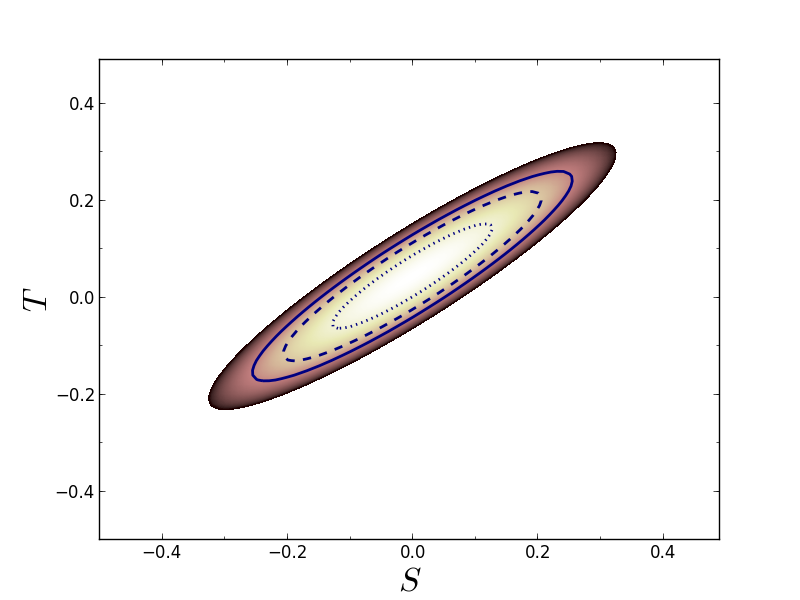}
\caption{\it Results of a $\chi^2$ analysis of $ST$ parameters in EWPTs using the expansion formalism of~\protect\cite{wellsandzhang}.
The dotted, dashed and solid contours denote the regions allowed at the $68\%, 95\%,$ and $99\%$ CL, respectively,
which may be compared with those of~\protect\cite{1209.2716}.}
\label{fig:ST}
\end{figure}

The $\hat{S}$ and $\hat{T}$ parameters are equivalent to a subset of the full set of dimension-6 operators
that can affect the EWPTs. In a redundant basis those entering in oblique corrections to vector boson self-energies are
\begin{equation*}
\mathcal{L}_\text{dim-6} \supset \frac{\bar{c}_{WB}}{m_W^2}\mathcal{O}_{WB} + \frac{\bar{c}_{W}}{m_W^2}\mathcal{O}_{W} + \frac{\bar{c}_{B}}{m_W^2}\mathcal{O}_{B} + \frac{\bar{c}_T}{v^2}\mathcal{O}_T + \frac{\bar{c}_{2W}}{m_W^2}\mathcal{O}_{2W} + \frac{\bar{c}_{2B}}{m_W^2}\mathcal{O}_{2B}	\, ,
\end{equation*}
while those that affect the leptonic and hadronic $Z$-pole measurements directly through modifications to the
gauge boson-fermion couplings are
\begin{equation*}
\mathcal{L}_\text{dim-6} \supset \sum_{f_L}\left( \frac{\bar{c}_{f_L}}{v^2}\mathcal{O}_{f_L} + \frac{\bar{c}^{(3)}_{f_L}}{v^2}\mathcal{O}^{(3)}_{f_L}\right) + \sum_{f_R}\frac{\bar{c}_{f_R}}{v^2}\mathcal{O}_{f_R}	\, .
\end{equation*}
The sum is over the left-handed lepton and quark doublets, $f_L \equiv L_L, Q_L$, 
and right-handed lepton and quark singlets, $f_R \equiv  e_R, u_R, d_R$, 
and we assume minimal flavour violation. The Fermi constant $G_F$ defined by the muon lifetime,
which we take as an input observable, is modified by $\bar{c}_{L_L}^{(3)l}$ via the four-fermion operators
$\mathcal{O}_{LL}^{(3)l}$:
\begin{equation*}
\mathcal{L}_\text{dim-6} \supset \frac{\bar{c}_{LL}^{(3)l}}{v^2}\mathcal{O}_{LL}^{(3)l}	\, .
\end{equation*}
We note that the coefficients are defined such that 
\begin{equation}
\bar{c} \equiv c\frac{M^2}{\Lambda^2}	\, ,
\label{eq:cbar}
\end{equation}
where $M \equiv v, m_W$ depending on the operator normalization, and $c \sim g_\text{NP}^2$ is a coefficient proportional to a new physics coupling $g_\text{NP}$ defined at the scale $M$. These are related to the coefficients at the new physics scale through RGE equations~\cite{RGEbounds}. 

These operators form a redundant basis that is reducible through field redefinitions, 
or equivalently the equations of motion, that have no effect on the S-matrix~\cite{eomreduction}. Following~\cite{PomarolRiva},
we may eliminate the operators $\mathcal{O}_{L_L}, \mathcal{O}^{(3)}_{L_L}$ that affect the left-handed
leptonic $Z$ couplings, and the operators $\mathcal{O}_{2W}, \mathcal{O}_{2B}, \mathcal{O}_{2G}$
corresponding to the $Y,W$ and $Z$ parameters~\cite{beyondSTU} in the generalization
of the universal oblique parameters~\footnote{The $U,V$ and $X$ parameters correspond to higher-dimensional operators.}.
The coefficients $\bar{c}_{WB}$ and the combination $\bar{c}_W + \bar{c}_B$ are related to the $\hat{S}$ parameter, and we eliminate the former using the identity
\begin{equation*}
\mathcal{O}_B = \mathcal{O}_{HB} + \frac{1}{4}\mathcal{O}_{BB} + \frac{1}{4}\mathcal{O}_{WB}	\, .
\end{equation*}
The operators $\mathcal{O}_{HB}, \mathcal{O}_{BB}$ affect Higgs physics and triple-gauge couplings,
as we shall see in the next section. Finally, the $\hat{T}$ parameter is equivalent to the $\bar{c}_T$ coefficient. These operators are listed in Table~\ref{tab:EWPToperators}.

The corrections to the self-energies are then as in (\ref{eq:selfenergiesST}),
with $\hat{S} = \bar{c}_W+\bar{c}_B$ and $\hat{T} = \bar{c}_T$. We also have the input observable correction
\begin{equation*}
\xi_{G_F} = -2\bar{c}^{(3)l}_{LL}	\, ,
\end{equation*}
and direct contributions to the output observables,
\begin{align*}
\xi_{\Gamma_Z} &= \frac{\Gamma_Z^l}{\Gamma_Z}\xi_{\Gamma_Z^l} + \frac{\Gamma_Z^\text{had}}{\Gamma_Z}\xi_{\Gamma_Z^\text{had}}	\, ,	\\
\xi_{\sigma^0_\text{had}} &= \xi_{\Gamma_Z^e} + \xi_{\Gamma_Z^\text{had}} - 2\xi_{\Gamma_Z}	\, , \\
\xi_{R_l} &= \xi_{\Gamma_Z^\text{had}} - \xi_{\Gamma_Z^l}	\, , \\
\xi_{R_q} &=  \xi_{\Gamma_Z^q} - \xi_{\Gamma_Z^\text{had}}	\, ,  \\
\xi_{A^{0,f}_\text{FB}} &= \xi_{A_e} + \xi_{A_f}	\, ,
\end{align*}
which can be written in terms of shifts to the $Z$-fermion couplings,
\begin{align*}
\xi_{A_f} &= \frac{4(g_Z^{f_L})^2(g_Z^{f_R})^2}{(g_Z^{f_L})^4 - (g_Z^{f_R})^4}\left( \xi_{g_Z^{f_L}} - \xi_{g_Z^{f_R}} \right)	\, , \\
\xi_{\Gamma_Z^f} &= \frac{2(g_Z^{f_L})^2}{(g_Z^{f_L})^2  + (g_Z^{f_R})^2}\xi_{g_Z^{f_L}} +  \frac{2(g_Z^{f_R})^2}{(g_Z^{f_L})^2  + (g_Z^{f_R})^2}\xi_{g_Z^{f_R}}	\, , 
\end{align*}
where
\begin{equation*}
\xi_{g_Z^{f_L}} = \frac{1}{{g_Z^{f_L}}}\left(T^3_f \bar{c}^{(3)}_{f_L} - \frac{\bar{c}_{f_L}}{2}\right)		\quad , \quad 
\xi_{g_Z^{f_R}} = -\frac{\bar{c}_{f_R}}{2g_Z^{f_R}}	\, ,
\end{equation*}
and $g_Z^f \equiv T^3_f - Q_f s^2_{\theta_W}$. 
Using these expressions and the expansion formalism in a $\chi^2$ analysis,
we obtain $95\%$ CL limits for the operator coefficients.

The left panel of Fig.~\ref{fig:EWPTsummary} shows our results for fits to the coefficients
$\bar{c}^{(3)l}_{LL}, \bar{c}_T, \bar{c}_W + \bar{c}_B$, together with the coefficient $\bar{c}^e_R$
that affects the leptonic observables 
$\{ \Gamma_Z, \sigma^0_\text{had}, R^0_e, R^0_\mu, R^0_\tau, A^{0,e}_\text{FB}, m_W \}$. 
The upper (green) bars indicate
the ranges for each of the coefficients varied individually, assuming that the other coefficients
vanish, and the lower (red) bars show the ranges for a global fit in which all the coefficients
are varied simultaneously. In both fits, the coefficients are all quite compatible with zero,
with ranges $\sim \pm 0.001$ in the single-coefficient analysis, increasing in the global fit
up to $\sim \pm 0.004$ for the coefficient $\bar{c}_T$ in the multi-coefficient analysis~\footnote{We note that
larger marginalized ranges for $\bar{c}^e_R$ and $\bar{c}^{(3)l}_{LL}$ are found in~\cite{PomarolRiva}, warranting further cross-checks.}. The legend at
the top of the left panel of Fig.~\ref{fig:EWPTsummary} translates the ranges of the coefficients
into ranges of sensitivity to a large mass scale $\Lambda$. We see that all the sensitivities are
in the multi-TeV range, including in the global analysis. 

\begin{figure}[h!]
\centering 
\includegraphics[scale=0.38]{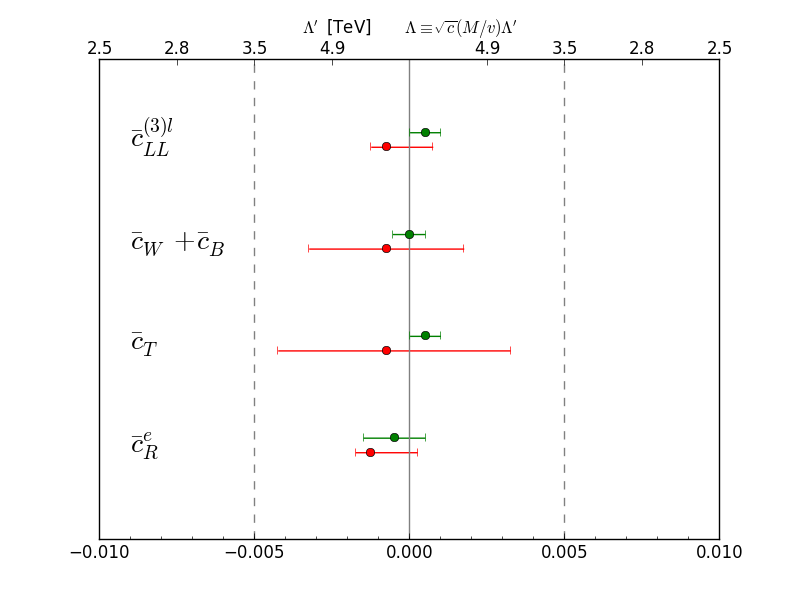}
\includegraphics[scale=0.38]{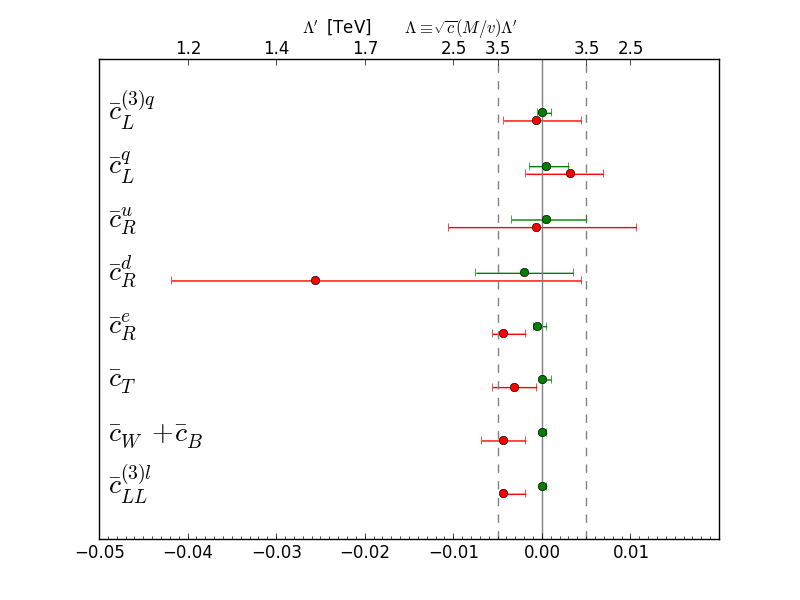}
\caption{\it The 95\% CL ranges found in analyses of the leptonic observables (left panel)
and including also the hadronic observables (right panel). In each case, the upper (green)
bars denote single-coefficient fits, and the lower (red) bars denote multi-coefficient fits. The upper-axis should be read $\times \frac{m_W}{v}\sim 1/3$ for $\bar{c}_W + \bar{c}_B$. }
\label{fig:EWPTsummary}
\end{figure}

The right panel of Fig.~\ref{fig:EWPTsummary} shows the effect of including the
hadronic observables, $\{R^0_b, R^0_c, A^{0,b}_\text{FB}, \\ A^{0,c}_\text{FB}, A_b, A_c\}$,
and the coefficients that contribute directly to them, namely $\bar{c}^q_L, \bar{c}^{(3)q}_L, \bar{c}^u_R$ and $\bar{c}^d_R$.     
The ranges for the single-variable fits
to $\bar{c}^{(3)l}_{LL}, \bar{c}_T, \bar{c}_W + \bar{c}_B$ and $\bar{c}^e_R$ (upper,green lines) are the same as in the left
panel, but the horizontal scales are different, as seen immediately by comparing the separations of the
vertical black dashed `tramlines'. The ranges of these coefficients are altered significantly in the
global 8-coefficient fit (lower, red lines) and we see significant tension with the null hypotheses for
$\bar{c}^{(3)l}_{LL}, \bar{c}_T, \bar{c}_W + \bar{c}_B$ and $\bar{c}^e_R$, which reflects the well-known
tension between the Standard Model and heavy-flavour measurements at the $Z$ peak. However,
values of $\bar{c}^{(3)l}_{LL}, \bar{c}_T, \bar{c}_W + \bar{c}_B$ and $\bar{c}^e_R$ between 0 and -0.01 are favoured,
corresponding to $\Lambda \gtrsim 2.5$~TeV. The ranges of $\bar{c}^q_L, \bar{c}^{(3)q}_L, \bar{c}^u_R$ and $\bar{c}^d_R$
are considerably broader in both fits, particularly in the global 8-coefficient fit, most notably $\bar{c}^u_R$ and $\bar{c}^d_R$,
with values of the latter approaching -0.05 being allowed at the 95\% CL.

\section{Triple-Gauge and Higgs Couplings at the LHC}

In previous work~\cite{ESY3} we used LHC measurements of Higgs signal strengths together with
differential distributions in Higgs associated production measurements by ATLAS and D0 to constrain
all the dimension-6 operators affecting Higgs physics. The associated production information was vital in
eliminating a blind direction, which can also be closed by including TGC measurements. These are most
precisely measured by LEP, but it has been recently pointed out that the LEP TGC constraints~\footnote{See
also \cite{trott} for a recent discussion on the use of TGC observables as reported by LEP for constraining 
dimension-6 operators in different bases.}
have a direction of limited sensitivity due to accidental partial cancellations~\cite{leshouches}. Meanwhile,
TGCs have been analysed at 8 TeV at the LHC by both the CMS and ATLAS experiments~\cite{CMS78TGC,ATLAS8TGC},
and here we study their potential to complement Higgs physics in constraining a complete set of dimension-6 operators.

\subsection{TGC Constraints on Dimension-6 Operator Coefficients}

\begin{table}[h]
\begin{center}
\begin{tabular}
{|c | c | c | c | c | c |}
\hline
\multirow{2}{*}{Operator} & \multirow{2}{*}{Coefficient} & \multicolumn{2}{| c |}{LHC Constraints}  \\
 &  & Individual & Marginalized \\
\hline
${\mathcal O}_W=\frac{ig}{2}\left( H^\dagger  \sigma^a \lra {D^\mu} H \right )D^\nu  W_{\mu \nu}^a$ & \multirow{2}{*}{$\frac{m_W^2}{\Lambda^2}(c_W - c_B)$} & \multirow{2}{*}{$(-0.022,0.004)$}  & \multirow{2}{*}{$(-0.035,0.005)$}   \\
${\mathcal O}_B=\frac{ig'}{2}\left( H^\dagger  \lra {D^\mu} H \right )\partial^\nu  B_{\mu \nu}$ &  &  &  \\
\hline
${\mathcal O}_{HW}=i g(D^\mu H)^\dagger\sigma^a(D^\nu H)W^a_{\mu\nu}$ & $\frac{m_W^2}{\Lambda^2}c_{HW}$ & $(-0.042,0.008)$ & $(-0.035,0.015)$  \\
\hline
${\mathcal O}_{HB}=i g^\prime(D^\mu H)^\dagger(D^\nu H)B_{\mu\nu}$ & $\frac{m_W^2}{\Lambda^2}c_{HB}$ & $(-0.053,0.044)$ & $(-0.045,0.075)$ \\
\hline
${\mathcal O}_{3W}= \frac{1}{3!} g\epsilon_{abc}W^{a\, \nu}_{\mu}W^{b}_{\nu\rho}W^{c\, \rho\mu}$ & $\frac{m_W^2}{\Lambda^2}c_{3W}$ & $(-0.083,0.045)$ & $(-0.083,0.045)$  \\
\hline
${\mathcal O}_{g}=g_s^2 |H|^2 G_{\mu\nu}^A G^{A\mu\nu}$ & $\frac{m_W^2}{\Lambda^2}c_{g}$ & $(0,3.0)\times 10^{-5}$ & $(-3.2,1.1)\times 10^{-4}$  \\
\hline
${\mathcal O}_{\gamma}={g}^{\prime 2} |H|^2 B_{\mu\nu}B^{\mu\nu}$ & $\frac{m_W^2}{\Lambda^2}c_{\gamma}$ & $(-4.0,2.3)\times 10^{-4}$ & $(-11,2.2)\times 10^{-4}$  \\
\hline
${\mathcal O}_H=\frac{1}{2}(\partial^\mu |H|^2)^2$ & $\frac{v^2}{\Lambda^2}c_{H}$ & $(-,-)$ & $(-,-)$  \\
\hline
${\mathcal O}_{f}   =y_f |H|^2    \bar{F}_L H^{(c)} f_R + \text{h.c.}$ & $\frac{v^2}{\Lambda^2}c_{f}$ & $(-,-)$ & $(-,-)$  \\
\hline
\end{tabular}
\end{center}
\caption{\it List of operators in our basis entering in LHC Higgs (including D0 associated production) and TGC physics, together with $95\%$ CL bounds when individual coefficients are switched on one at a time, and marginalized in a simultaneous fit.}
\label{tab:LHCoperators}
\end{table}

The operators affecting Higgs physics and TGCs in the basis we adopt are listed in Table~\ref{tab:LHCoperators}, with the Lagrangian given by
\begin{align*}
\mathcal{L}_\text{dim-6} & \supset \frac{\bar{c}_{W}}{m_W^2}\mathcal{O}_{W} + \frac{\bar{c}_{B}}{m_W^2}\mathcal{O}_{B} + \frac{\bar{c}_{HW}}{m_W^2}\mathcal{O}_{HW} +  \frac{\bar{c}_{HB}}{m_W^2}\mathcal{O}_{HB} + \frac{\bar{c}_\gamma}{m_W^2}\mathcal{O}_{\gamma} + \frac{\bar{c}_g}{m_W^2}\mathcal{O}_g 		\\
& + \frac{\bar{c}_{3W}}{m_W^2}\mathcal{O}_{3W} + \sum_{f=t,b,\tau} \frac{\bar{c}_f}{v^2}\mathcal{O}_f + \frac{\bar{c}_H}{v^2}\mathcal{O}_H + \frac{\bar{c}_6}{v^2}\mathcal{O}_6	\, . 
\end{align*}
The constraint at the per-mille level on the combination $\bar{c}_W + \bar{c}_B$ obtained in the previous
Section allows us to set $\bar{c}_B = -\bar{c}_W$ (or equivalently to constrain the direction $\bar{c}_W - \bar{c}_B$). Ignoring the unconstrained operator $\mathcal{O}_6$
that affects the Higgs self-couplings and (for simplicity) setting $\bar{c}_b = \bar{c}_\tau \equiv \bar{c}_d$ then
reduces the number of independent coefficients to nine. The coefficients $\bar{c}_W, \bar{c}_{HW}, \bar{c}_{HB}$
and $\bar{c}_{3W}$ affect TGCs, with $\bar{c}_{3W}$ being limited only by TGC measurements,
since it does not affect Higgs physics. 

\begin{figure}[h!]
\centering
\includegraphics[scale=0.4]{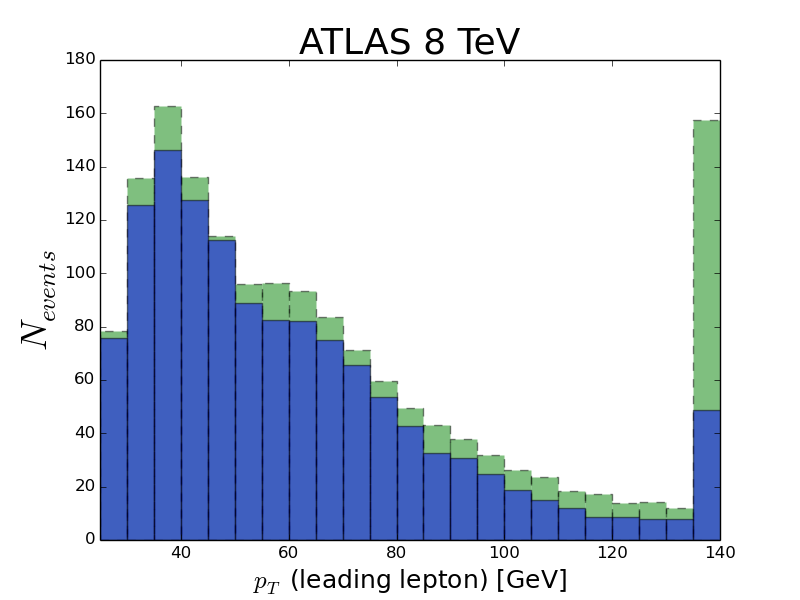}
\caption{\it The same-flavour $p_T$ distribution of the leading lepton after the TGC analysis
cuts for ATLAS at 8 TeV. The Standard Model distribution is shown in blue with solid lines,
and the effect of $\bar{c}_{HW} = 0.1$ is superimposed in green with dashed lines. }
\label{fig:exampledistribution}
\end{figure}

\begin{figure}[h!]
\centering
\includegraphics[scale=0.35]{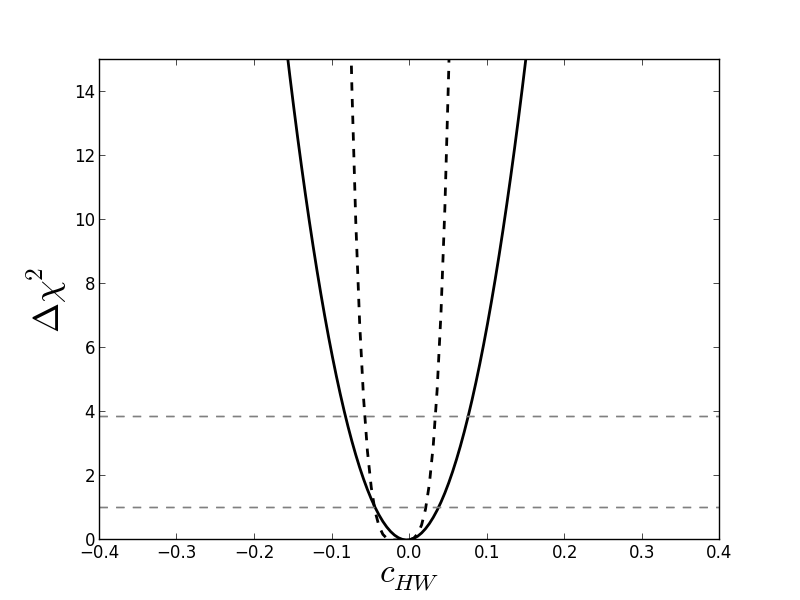}
\includegraphics[scale=0.35]{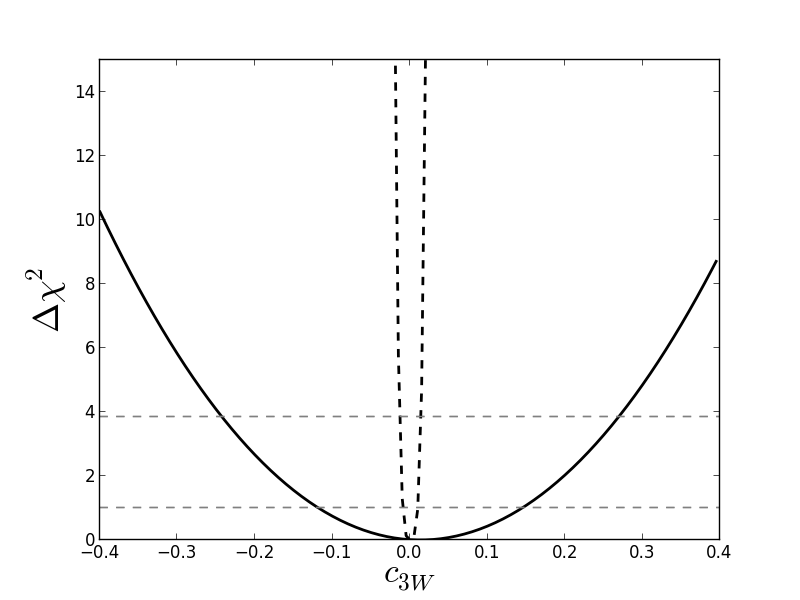}
\caption{\it Comparisons between the $\chi^2$ functions from fits to the same-flavour ATLAS distribution including only linear (solid lines) and
also quadratic (dashed lines) dependences on the dimension-6 coefficients $\bar{c}_{HW}$ (left panel)
and $\bar{c}_{3W}$ (right panel).}
\label{fig:quadvslinear}
\end{figure}

We calculate the TGCs in the presence of dimension-6 operators using the {\tt FeynRules}
implementation of~\cite{benj} in {\tt MadGraph v2.1.2}~\cite{mg5}, interfaced with {\tt Pythia}~\cite{pythia}
and {\tt Delphes}~\cite{delphes}. In the case of ATLAS, we implement the analysis given in~\cite{ATLAS8TGC}.
This requires events that pass the selection cuts to have exactly 2 opposite-sign leptons with no jets,
$p_T > 25 (20)$ GeV for leading (sub-leading) leptons, $m_{ll} > 15 (10)$ GeV and $E_T^\text{miss} > 45 (15)$ GeV
for same-flavour (different-flavour) lepton pairs, as well as $|m_{ll} - m_Z| > 15$ GeV for the same-flavour case.
Similarly, following~\cite{CMS78TGC}, for the CMS cuts we require 2 opposite-sign leptons with $p_T > 20$ GeV, 
total lepton $p_T > 45$ GeV and $75$~GeV $< m_{ll} < 105$~GeV, $E_T^\text{miss} > 37 (20)$~GeV and
$m_{ll} > 20 (12)$ GeV for same-flavour (opposite-flavour) pairs, and no jets with $|\eta| < 5, E_T > 30$ GeV.

The resulting $p_T$ distribution of the leading lepton for the ATLAS 8~TeV analysis is shown in Fig.~\ref{fig:exampledistribution}
including $\bar{c}_{HW} = 0.1$ as well as the Standard Model contribution~\footnote{The applicability of
the effective field theory approach to this TGC analysis is discussed in the Appendix.}. We focus on the number of events in the last
(overflow) bin, since this has the highest signal-to-background ratio and grows rapidly as a function of this and
the other dimension-6 coefficients~\footnote{The validity of the effective field theory at such high $p_T$ may be
restricted only to certain models~\cite{knocheletal1406.7320}, but the range of validity will increase as the
current precision of LHC TGC measurements is improved.}. We prefer to keep only the linear dependences
on the dimension-6 coefficients, considering that
it is not consistent to keep terms that are quadratic in the dimension-6 coefficients if one does not have
reason to expect that the coefficients of dimension-8 operators would be suppressed.
As an example, we note that the signal-strength dependence of the overflow bin on $\bar{c}_{HW}$ for the ATLAS 8-TeV same-flavour distribution is found to be 
\begin{equation*}
\mu_\text{last-bin}^\text{ATLAS8} = 1 + 3.45 \bar{c}_{HW} + 234 \bar{c}_{HW}^2	\, ,
\end{equation*}
and we keep only the linear term in our global fits.
The constraints obtained using this linear (quadratic) dependence on the dimension-6 coefficients
are plotted as solid (dashed) lines in Fig.~\ref{fig:quadvslinear}. The left panel is for $\bar{c}_{HW}$,
and right panel is for $\bar{c}_{3W}$. When deriving constraints we use the background and Standard Model signal Monte-Carlo (MC) distributions
of the leading lepton $p_T$ provided by the experiments, and marginalize over the MC error. This is given along with the observed number of 
events and their errors in \cite{ATLAS8TGC} for ATLAS and \cite{CMS78TGC} for CMS. We see that the quadratic and linear fits for $\bar{c}_{HW}$ are quite similar, whereas
the constraint from the (preferred) linear fit for $\bar{c}_{3W}$ is significantly weaker than that from the (deprecated) quadratic fit.

\begin{figure}[h!]
\centering
\includegraphics[scale=0.25]{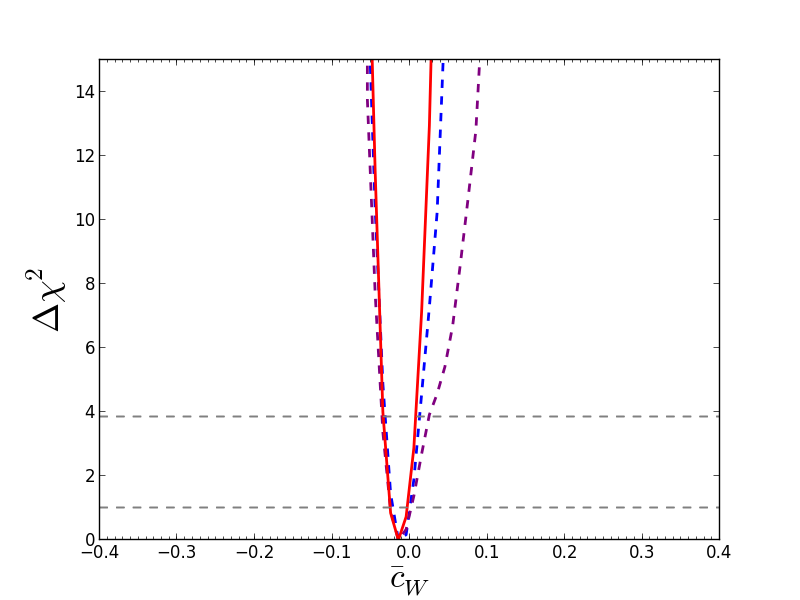}
\includegraphics[scale=0.25]{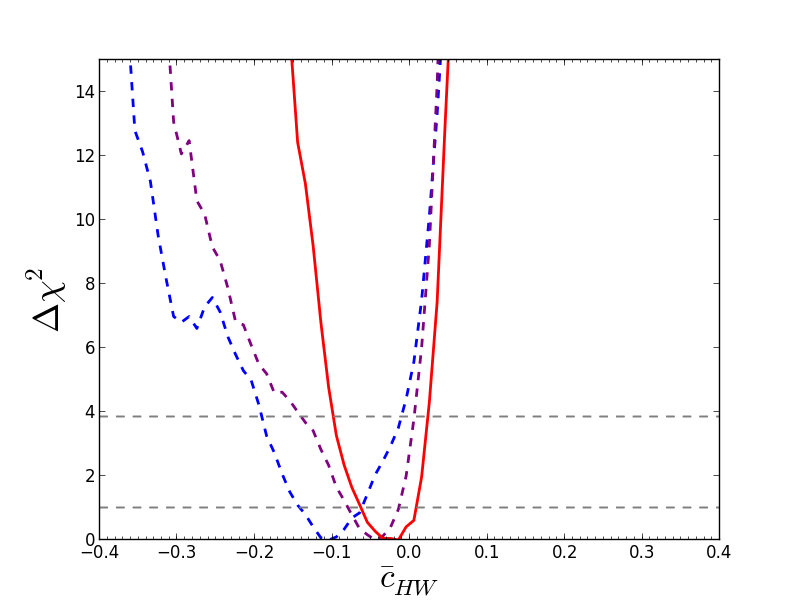}
\includegraphics[scale=0.25]{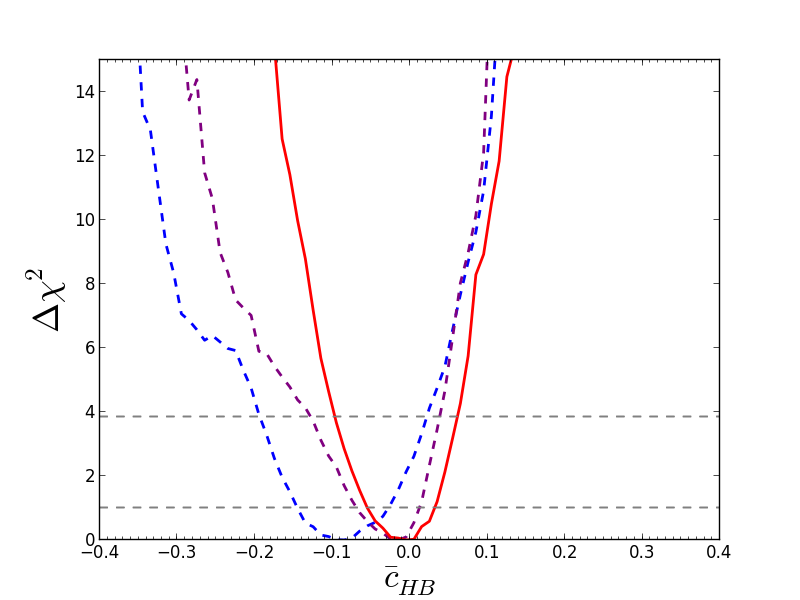} \\
\includegraphics[scale=0.25]{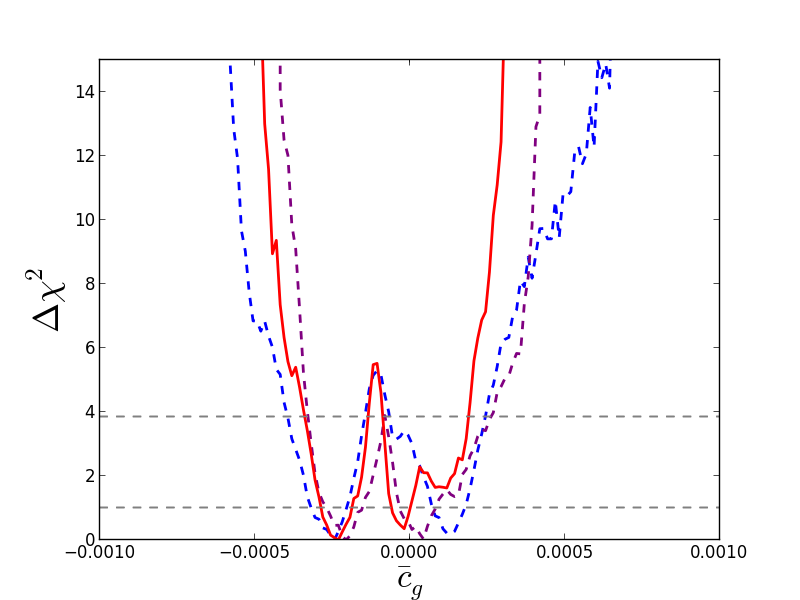}
\includegraphics[scale=0.25]{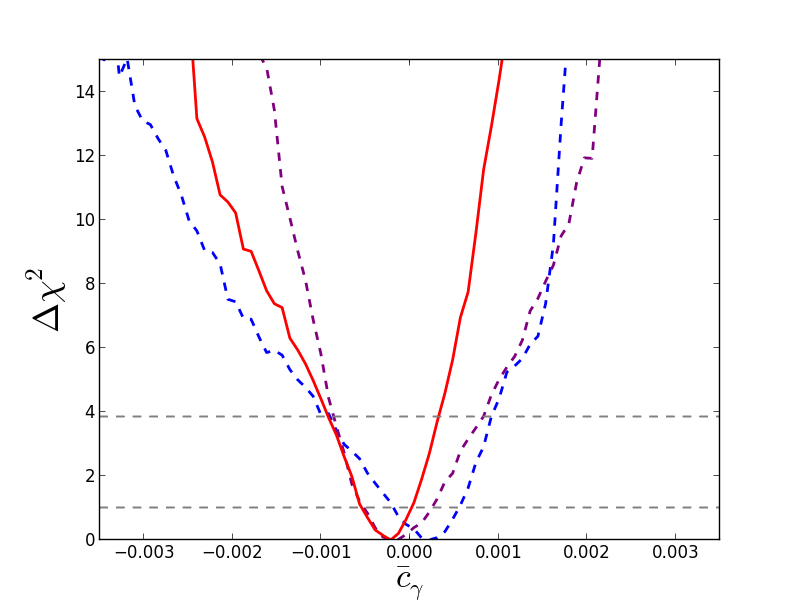}
\includegraphics[scale=0.25]{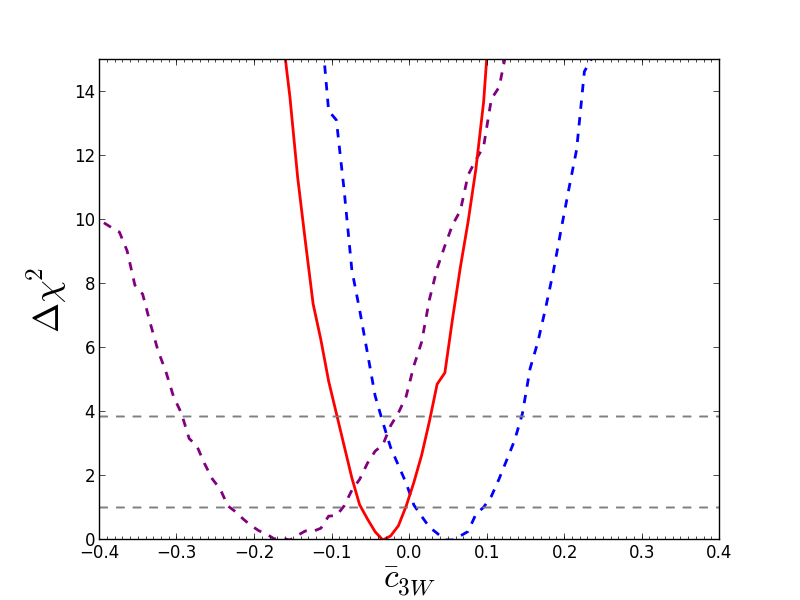} \\
\includegraphics[scale=0.25]{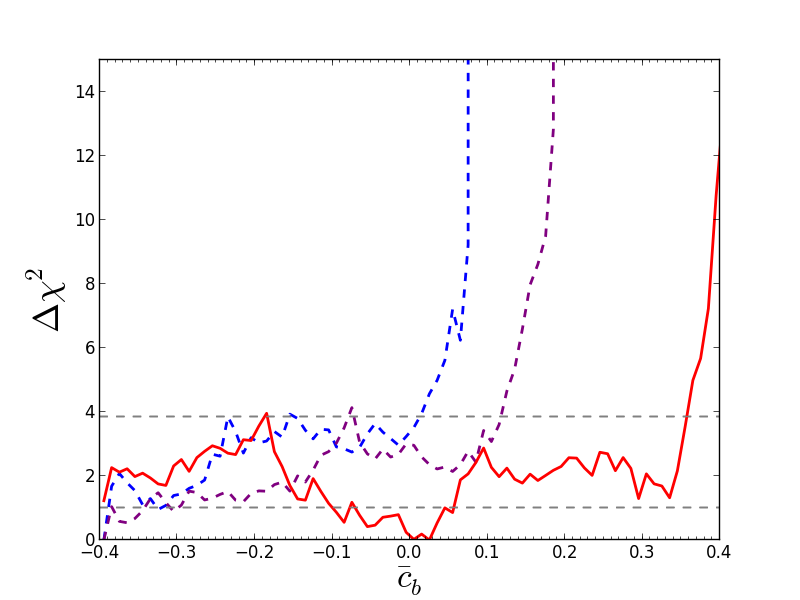}
\includegraphics[scale=0.25]{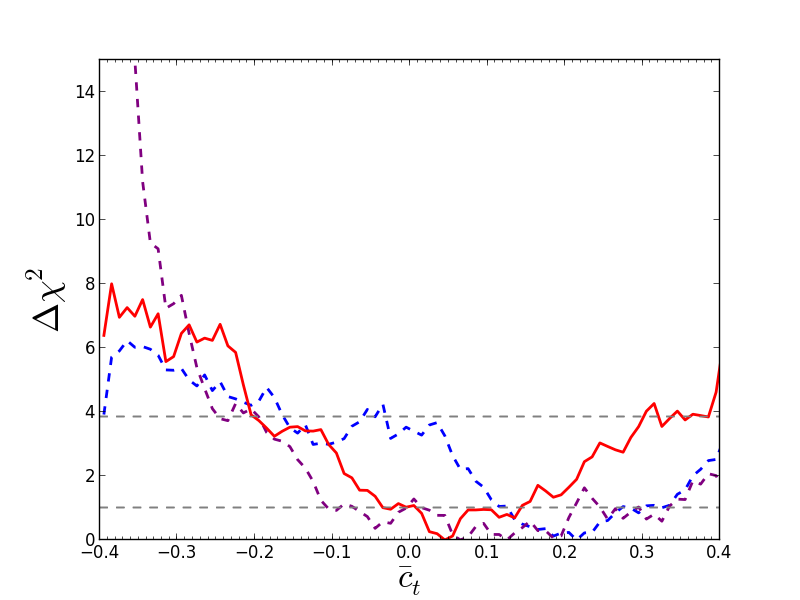}
\includegraphics[scale=0.25]{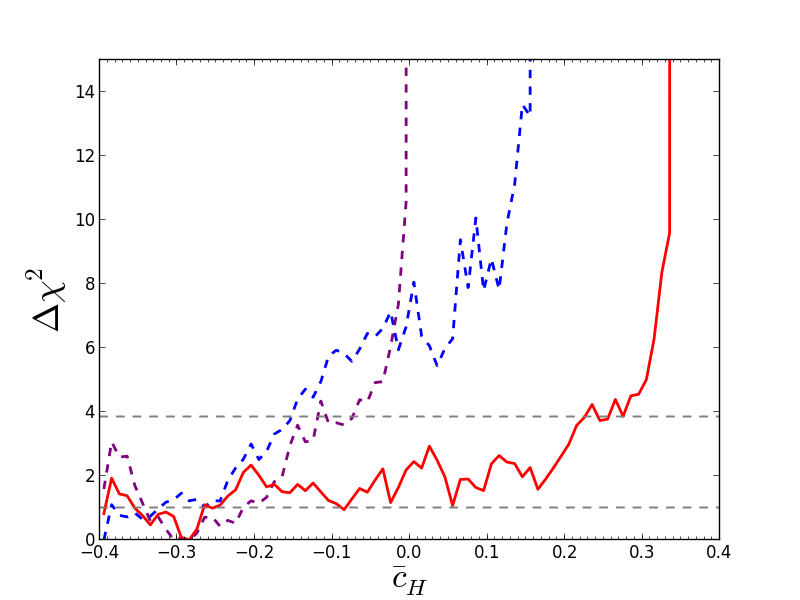} \\
\caption{\it Comparisons of the constraints on the dimension-6 coefficients 
$\bar{c}_{W}$, $\bar{c}_{HW}$ and $\bar{c}_{HB}$ (top row),
$\bar{c}_{g}$, $\bar{c}_{\gamma}$ and $\bar{c}_{3W}$ (middle row), and
$\bar{c}_{b}$, $\bar{c}_{t}$ and $\bar{c}_{H}$ (bottom row) provided by the LHC signal-strength
data together with the ATLAS 8-TeV (purple lines), the CMS
7- and 8-TeV TGC measurements (blue lines) and their combination (red lines).}
\label{fig:ATLASCMS}
\end{figure}

For the full global fit we use the same-flavour and different-flavour distributions for ATLAS at 8 TeV and the CMS 7 and 8 TeV data. In Fig.~\ref{fig:ATLASCMS} we compare the constraints from the combination of
the ATLAS and CMS TGC measurements with the LHC Higgs signal-strength data
on each of the dimension-6 coefficients $\bar{c}_{W}$, $\bar{c}_{HW}$ and $\bar{c}_{HB}$ (top row),
$\bar{c}_{g}$, $\bar{c}_{\gamma}$ and $\bar{c}_{3W}$ (middle row), and
$\bar{c}_{b}$, $\bar{c}_{t}$ and $\bar{c}_{H}$ (bottom row)~\footnote{We note that the constraints
on the last three operators are relatively weak, but include them for information.}
The purple line represents the combination
of LHC signal-strength constraints with the ATLAS 8-TeV TGC measurements, the blue line the 
combination of CMS 7- and 8-TeV constraints, and the red line uses all the sets of LHC TGC constraints.
We use the signal-strength information on the $W^+{W^-}^{(*)}, ZZ^{(*)},\gamma\gamma, Z\gamma$, and $\tau^+\tau^-$ final states,
whose likelihoods are obtained as explained in~\cite{ESY3}. We observe that the constraints on the coefficient $\bar{c}_{3W}$, which only affects TGCs, is at the same level as some of the other coefficients whose operators also affect Higgs physics. 

\begin{figure}[h!]
\centering
\includegraphics[scale=0.5]{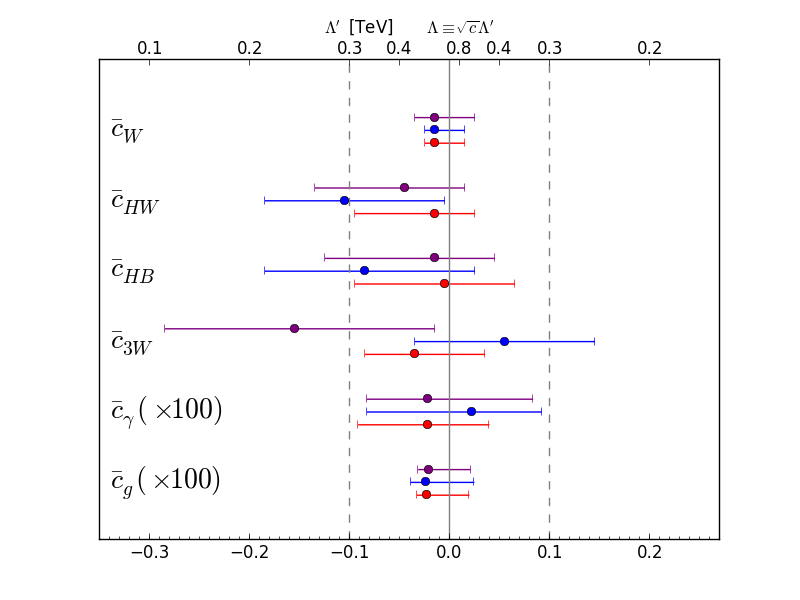}
\caption{\it The marginalised 95\% CL ranges for the dimension-6 operator coefficients obtained by
combining the LHC signal-strength data with the ATLAS 8-TeV TGC data (purple bars), the CMS
7- and 8-TeV TGC measurements (blue bars), and their combination (red bars). Note that $\bar{c}_{\gamma,g}$ are shown $\times 100$, so for these coefficients the upper axis should therefore be read $\times 10$.}
\label{fig:ATLASCMSsummary}
\end{figure}

The results in Fig.~\ref{fig:ATLASCMS} are summarised in the marginalised 95\% CL  ranges
displayed in Fig.~\ref{fig:ATLASCMSsummary}. Again, the LHC signal-strength data
are always included, in combination with the ATLAS 8-TeV data (purple bars), the
CMS 7- and 8-TeV data (blue bars) and all the LHC TGC data (red bars). As already
mentioned, the LHC TGC data enables a competitive model-independent bound on the coefficient $\bar{c}_{3W}$.

\subsection{Inclusion of Higgs Associated Production Constraints}

\begin{figure}[h!]
\centering
\includegraphics[scale=0.25]{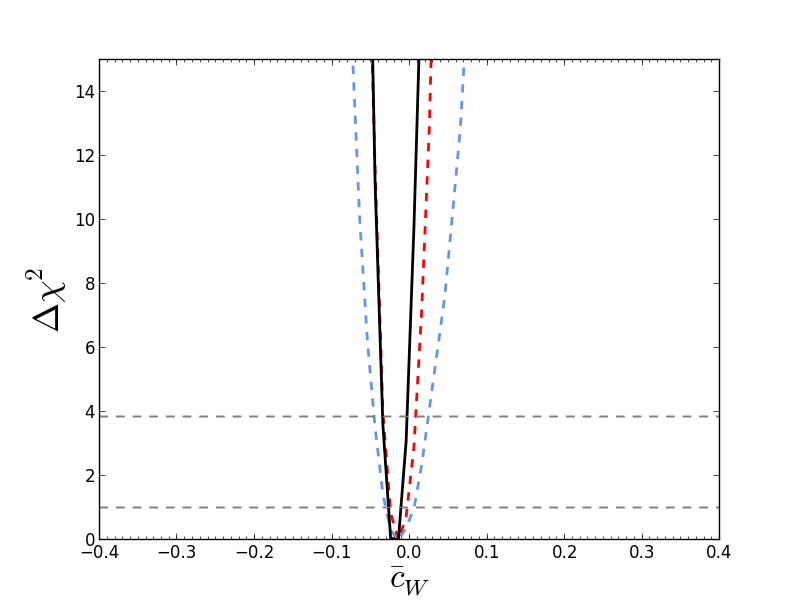}
\includegraphics[scale=0.25]{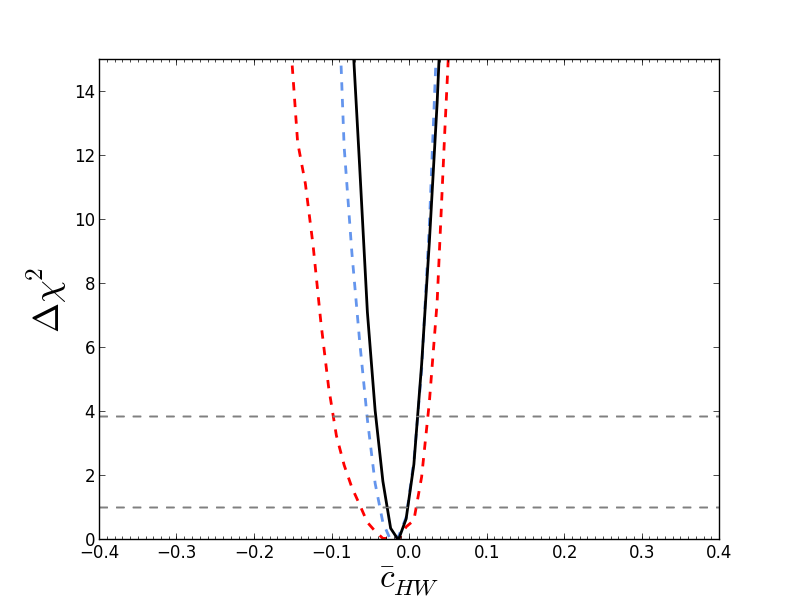}
\includegraphics[scale=0.25]{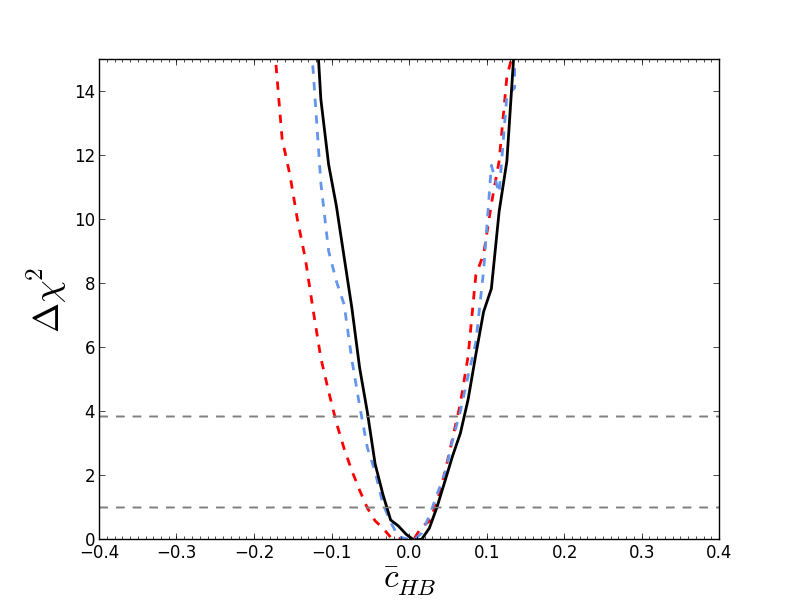} \\
\includegraphics[scale=0.25]{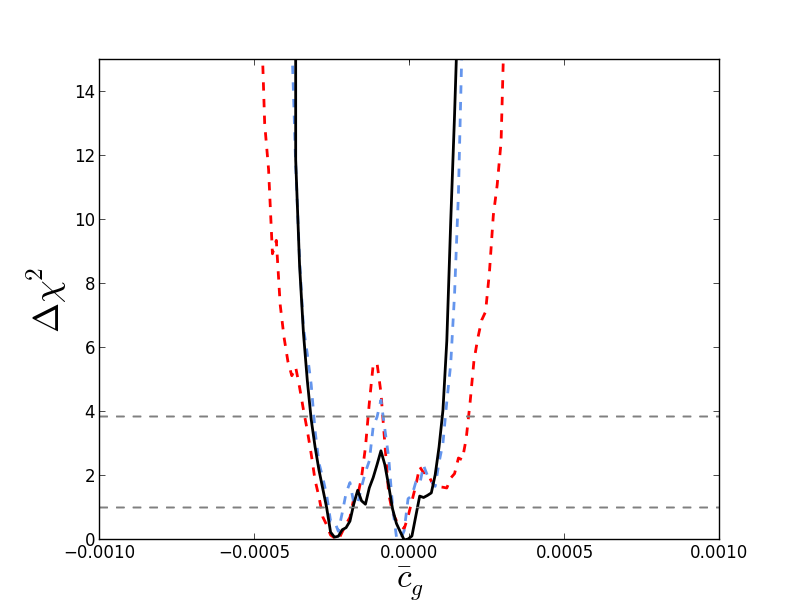}
\includegraphics[scale=0.25]{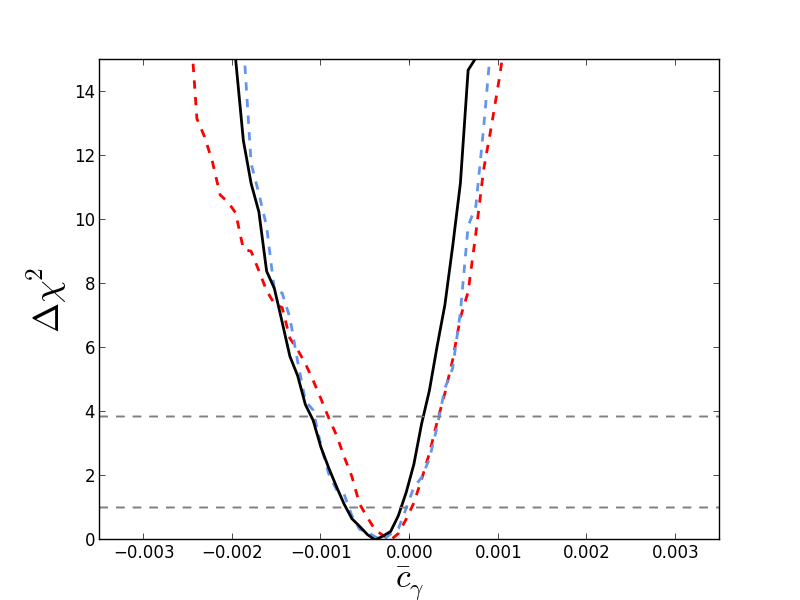}
\includegraphics[scale=0.25]{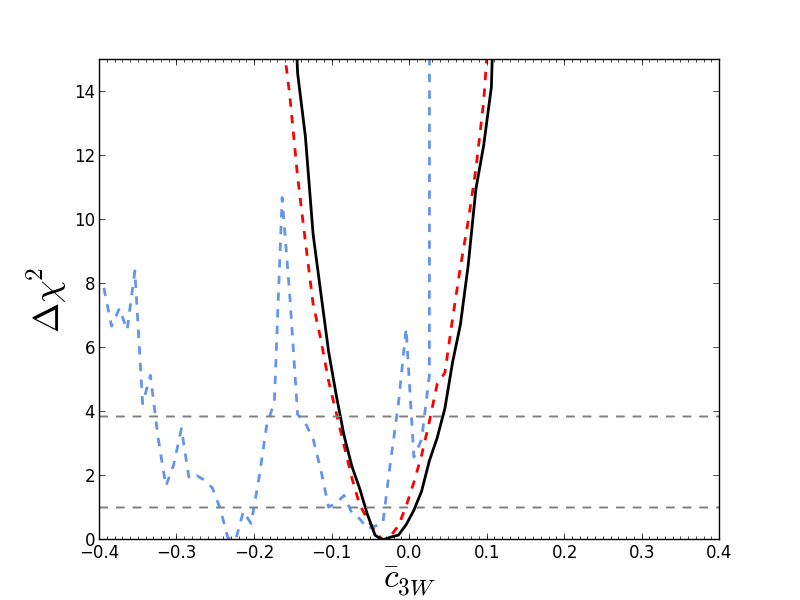} \\
\includegraphics[scale=0.25]{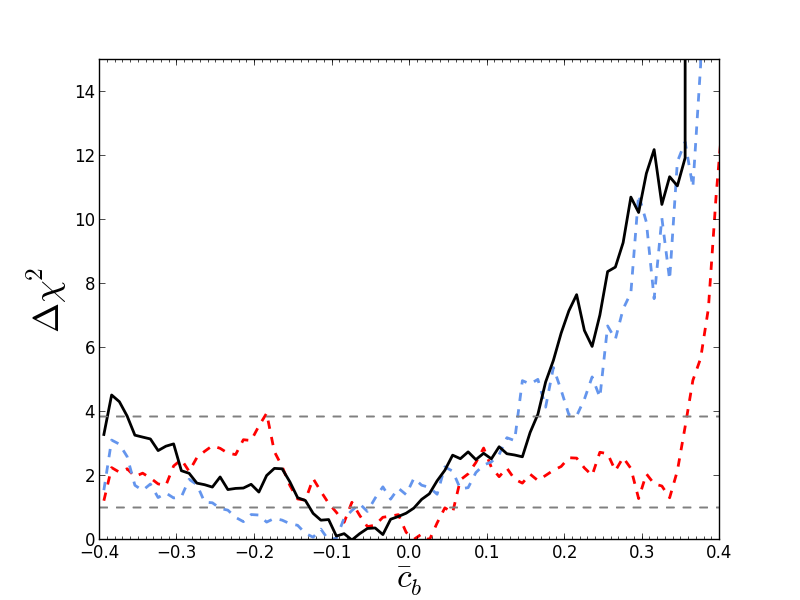}
\includegraphics[scale=0.25]{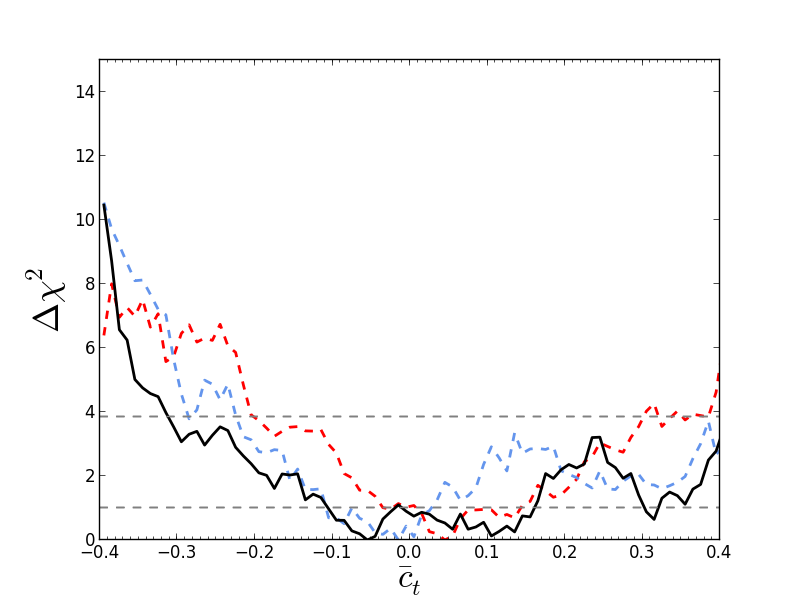}
\includegraphics[scale=0.25]{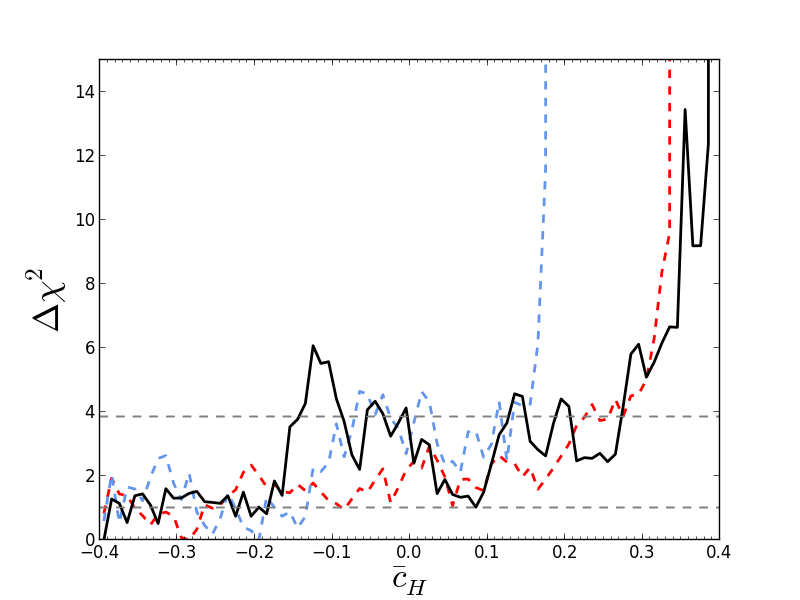} \\
\caption{\it The marginalised $\chi^2$ distributions for each of the
dimension-6 coefficients $\bar{c}_{W}$, $\bar{c}_{HW}$ and $\bar{c}_{HB}$ (top row),
$\bar{c}_{g}$, $\bar{c}_{\gamma}$ and $\bar{c}_{3W}$ (middle row), and
$\bar{c}_{b}$, $\bar{c}_{t}$ and $\bar{c}_{H}$ (bottom row),
including the signal strengths measured at the LHC and the constraints from the kinematic distributions for
associated $H + V$ production measured by ATLAS and D0 (dashed blue lines),
the signal strengths and the LHC TGC measurements (red lines), and all the constraints (black lines).}
\label{fig:marginalizedchisquared}
\end{figure}

We now include in our analysis the constraints from the kinematics of associated Higgs production,
following the analysis of~\cite{ESY3}~\footnote{The applicability of
the effective field theory approach to this associated production analysis is discussed in the Appendix.}.
Fig.~\ref{fig:marginalizedchisquared} displays the marginalised $\chi^2$ distributions for each of the
dimension-6 coefficients $\bar{c}_{W}$, $\bar{c}_{HW}$ and $\bar{c}_{HB}$ (top row),
$\bar{c}_{g}$, $\bar{c}_{\gamma}$ and $\bar{c}_{3W}$ (middle row), and
$\bar{c}_{b}$, $\bar{c}_{t}$ and $\bar{c}_{H}$ (bottom row)~\footnote{We note again that the constraints
on the last three operators are relatively weak, but include them for information.}.
In each panel, the dashed blue  line includes the Higgs signal strengths measured at the LHC and the
constraints from the kinematic distributions for associated $H + V$ production measured by ATLAS and D0,
whereas the solid red line includes the signal strengths and the LHC TGC measurements. The solid
black lines include all the constraints: the signal strengths, the kinematic distributions and the TGCs measured at the LHC.
We see that the LHC TGC measurements are the strongest for $\bar{c}_{W}$ and $\bar{c}_{3W}$: in
particular, they are necessary to obtain any meaningful constraint on $\bar{c}_{3W}$, which is not
constrained at all by Higgs physics alone. On the other hand, the Higgs constraints are more
important for $\bar{c}_{HW}$, $\bar{c}_{HB}$ and $\bar{c}_{g}$, whereas the TGC and Higgs constraints
are of comparable importance for the other coefficients.

\begin{figure}[h!]
\centering
\includegraphics[scale=0.5]{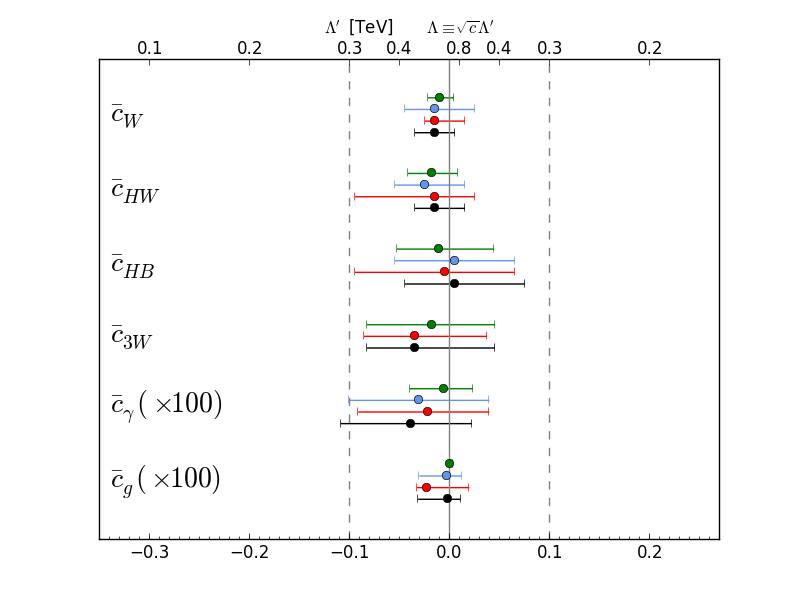}
\caption{\it The 95\% CL constraints obtained for single-coefficient fits (green bars),
and the marginalised 95\% ranges for the
LHC signal-strength data combined with the kinematic distributions for associated $H + V$ production
measured by ATLAS and D0 (blue bars), combined with the LHC TGC data (red lines), and the global combination with
both the associated production and TGC data (black bars). Note that $\bar{c}_{\gamma,g}$ are shown $\times 100$, so for these coefficients the upper axis should therefore be read $\times 10$.}
\label{fig:fitsummary}
\end{figure}

The results of our fits are summarised in Fig.~\ref{fig:fitsummary}. The individual 95\% CL constraints obtained by switching 
one coefficient on at a time are shown as green bars. The other lines are the marginalised 95\% ranges obtained using the
LHC signal-strength data in combination with the kinematic distributions for associated $H + V$ production
measured by ATLAS and D0 (blue bars), in combination with the LHC TGC data (red lines), and in combination with
both the associated production and TGC data (black bars). We see again that the LHC TGC constraints
are the most important for $\bar{c}_{W}$ and $\bar{c}_{3W}$, whereas the Higgs constraints are more
important for $\bar{c}_{HW}$, $\bar{c}_{HB}$ and $\bar{c}_{g}$. Our numerical results for the 95\% CL ranges for these coefficients are shown alongside the operator definitions in Table~\ref{tab:LHCoperators}.

\section{Application to the Two-Higgs Doublet Model}

We now discuss an example of the application of our constraints to a specific
ultra-violet (UV) completion of the effective field theory. The case of a singlet scalar and stops contributing to dimension-6 operators was recently considered in \cite{precisionhiggs}. 
Here we briefly look at applying our constraints to the 2HDM scenario, which is worth further investigation~\cite{inprogress}.

In a large range of models, the only coupling of the
Higgs to massive vector bosons has the following Lorentz structure
\bear
h W_{\mu\nu} W^{\mu\nu}   \ .
\eear
The translation between this Higgs anomalous coupling and the operators is given
in~\cite{benj} (see also~\cite{masso}). The following constraints
\begin{equation}
{\bar c}_{HW} \; =-\; {\bar c}_W, \; \; {\bar c}_{HB} \; = \; - {\bar c}_B
\label{2HDM}
\end{equation}
are then satisfied at the UV scale.
We recall from Section~2 that, in addition, the EWPTs impose the constraint ${\bar c}_W \simeq - {\bar c}_B$,
implying that, to a good approximation
\begin{equation}
{\bar c}_W \; = \; - {\bar c}_B \; = \; - {\bar c}_{HW} \; = \; {\bar c}_{HB}  \, ,
\label{alsoEWPTs}
\end{equation}
with corrections due to renormalization-group running effects
that are negligible compared to the precision of the current LHC constraints. Moreover, in the 2DHM one also finds generically that ${\bar c}_{3W}$ is suppressed~\cite{inprogress}
\begin{equation}
{\bar c}_{3W}  \; \sim \; {\cal O} (0.1) g^2 {\bar c}_{HW}  \, , 
\label{smallc3W}
\end{equation}
so that it can be an order of magnitude smaller. In our application to the 2HDM we set it to zero,
as well as using the constraints (\ref{alsoEWPTs}). 

Examples of models in this class include a general two-Higgs doublet model (2HDM)~\cite{inprogress},
supersymmetry with electroweakino/sfermion loops~\cite{hollik},
and the exchange of a radion/dilaton particle~\cite{masso}. In the former two models these operators are
generated at loop level, whereas in the third case the operators appear at tree-level through the exchange of the
radion/dilaton particle. In the loop-induced cases, the validity of the effective theory is typically $\sqrt{\hat s} \sim 2 M$,
where $M$ is the mass scale of the heavy states.

Fig.~\ref{fig:2HDM} shows the $\chi^2$ distributions we find in a global fit to the three
independent dimension-6 coefficients of the 2HDM, ${\bar c}_W$, $\bar{c}_{g}$
and $\bar{c}_{\gamma}$ obtained under these assumptions. These distributions
have been obtained including all the constraints from the signal strengths measured at the LHC, 
the constraints from the kinematic distributions for
associated $H + V$ production measured by ATLAS and D0, and the LHC TGC measurements.
We find the following 95\% CL ranges
\begin{eqnarray}
{\bar c}_W & \in & - (0.02, 0.0004) \nonumber \\
{\bar c}_g & \in & - (0.00004, 0.000003) \nonumber \\
{\bar c}_\gamma & \in & - (0.0006, - 0.00003)
\label{2HDM95}
\end{eqnarray}
in this particular class of models.

\begin{figure}[h!]
\centering
\includegraphics[scale=0.25]{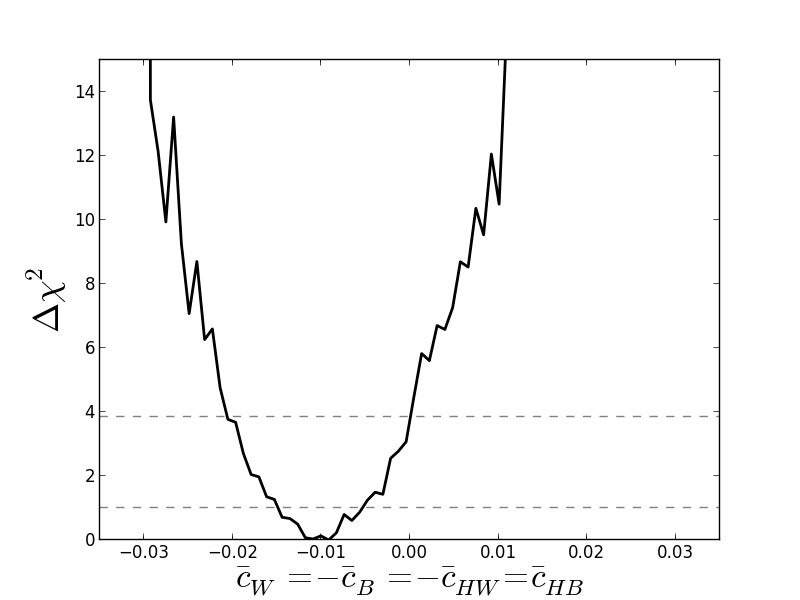}
\includegraphics[scale=0.25]{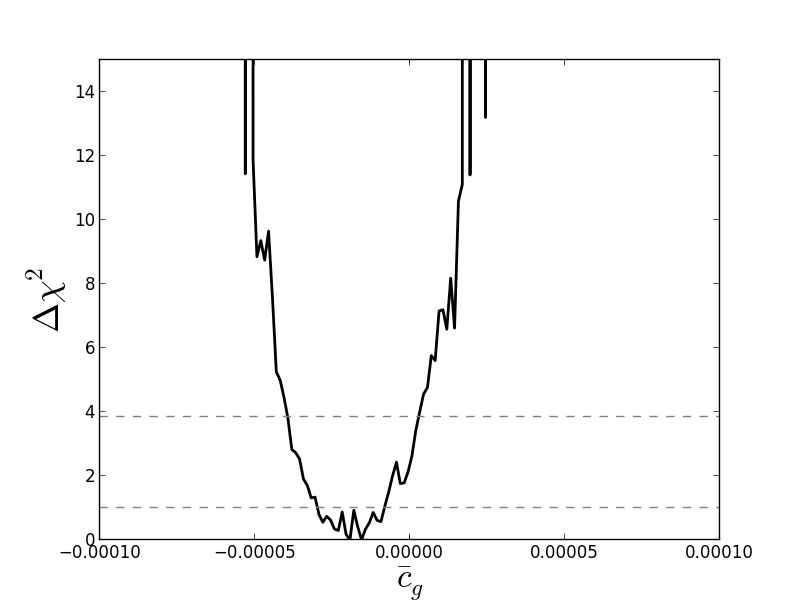}
\includegraphics[scale=0.25]{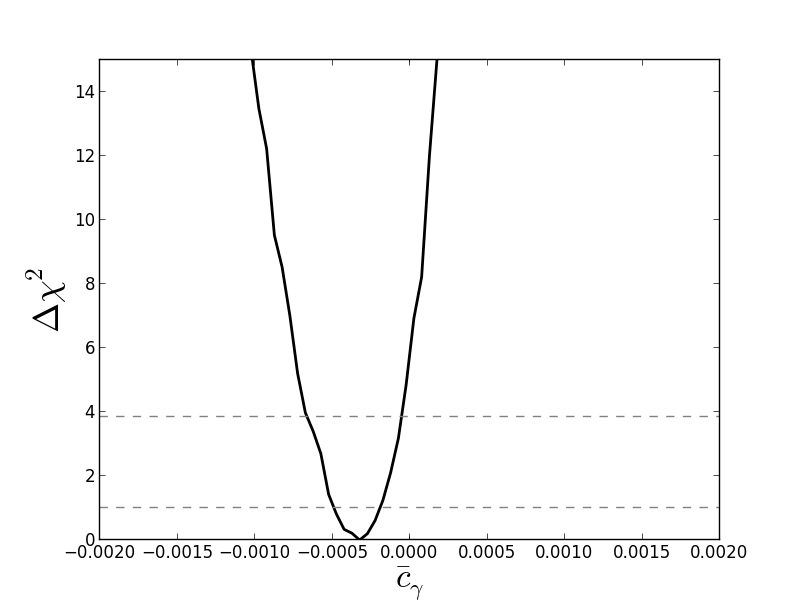} \\
\caption{\it The marginalised $\chi^2$ distributions for the coefficients
$\bar{c}_{W} = - \bar{c}_B = -\bar{c}_{HW} = \bar{c}_{HB}$, $\bar{c}_{g}$, and $\bar{c}_{\gamma}$ 
of the three independent dimension-6 operators in the 2HDM under the assumptions stated in the text.}
\label{fig:2HDM}
\end{figure}

\section{Conclusions}

The main lesson learned from Run I of the LHC is that, to a first approximation,
we seem to have a Standard Model-like Higgs sector. Taken together with the fact that there is currently no clear
evidence for any new physics beyond the Standard Model, it is natural to consider the Standard Model in its
complete effective theory formulation. Such a (relatively) model-independent framework 
parameterises all the possible ways in which decoupled new physics may affect
measurements at different experiments in a correlated and motivated way.  

We have analysed in this paper the constraints imposed on the coefficients of dimension-6
operator extensions of the Standard Model by EWPTs and LHC data.
We first analysed the EWPTs using the expansion formalism of~\cite{wellsandzhang},
which is particularly appropriate for models where the dominant corrections to the Standard Model
predictions are not necessarily present only in the vector-boson self-energies, as is the case for
general dimension-6 extensions of the Standard Model. We confirm previous findings that the EWPTs
provide particularly important constraints on some of the operator coefficients, as shown
in Fig.~\ref{fig:EWPTsummary} and Table~\ref{tab:EWPToperators}.

We then analysed the TGC data now available from ATLAS at 8 TeV and from CMS at 7 and 8 TeV.
We find that the most important aspects of the data are the highest-energy (overflow) bins in the lepton $p_T$
distributions, as illustrated in Fig.~\ref{fig:exampledistribution}, and use these together with
Higgs signal strength measurements to obtain constraints on a set of nine operator coefficients,
as shown in Figs.~\ref{fig:ATLASCMS} and \ref{fig:ATLASCMSsummary}. We then combined these LHC TGC constraints
with the constraints provided by measurements of the kinematics of Higgs production in
association with massive vector bosons at the Tevatron and the LHC, obtaining the results shown in
Figs.~\ref{fig:marginalizedchisquared} and \ref{fig:fitsummary} and Table~\ref{tab:LHCoperators}. As seen there, we find that completing the
Higgs signal strengths constraints on dimension-6 operators using the LHC TGCs provide the strongest
LHC constraints on some of the coefficients, whereas the Higgs differential distributions in associated production are more important for some others, with both making important contributions in some cases. In particular, we obtain the first bounds on the coefficient $\bar{c}_{3W}$ for a complete basis in the effective Standard Model. It is only by combining the TGC and Higgs constraints that one can obtain a complete picture of the
possible ranges of the dimension-6 operator coefficients after LHC Run~1.

It is to be expected that Run 2 of the LHC will provide important improvements in the sensitivity of
LHC probes of possible dimension-6 operators. These improvements will come not only from the
greater statistics, but also from the greater kinematic range that will strengthen the power of the
associated Higgs production kinematics and the TGC constraints, in particular. At the moment we
know that the Standard Model is very effective: LHC Run~2 data will give us a better idea just how effective
it is, and perhaps provide some pointers to the nature of the new physics that surely lies beyond it at
higher energies.

\section*{Acknowledgements}

We thank Francesco Riva for useful conversations and Maxime Gouzevitch and Alexander Savin
for helpful information about the CMS TGC distributions. The work of JE was supported partly by the London Centre
for Terauniverse Studies (LCTS), using funding from the European Research Council via the Advanced Investigator
Grant 26732, and partly by the STFC Grant ST/J002798/1.
The work of VS was supported by the STFC Grant ST/J000477/1. The work of TY was supported by a
Graduate Teaching Assistantship from King's College London.

\section*{Appendix: Kinematics and the Validity of the Effective Field Theory}

We use in Section 3 triple-gauge couplings and information on kinematic distributions in
Higgs production in association with a vector boson production constraints,
finding that typical 95\% CL constraints on the dimension-6 coefficients are ${\cal O}(10^{-1}-10^{-2})$.
For example, for the operator $\bar c_W$ our limits are
\bear
\bar c_{W} \in (-0.022,0.004) \textrm{ [one-by-one] and } (-0.035,0.005) \textrm{ [global]} \, .
\eear
Recalling the definition of the barred coefficients in Eq.~\ref{eq:cbar}, one can interpret these limits in terms of new physics at scale $\Lambda$ coupled to the SM with
strength $g_\text{NP}$, 
\bear
 \frac{\bar c_{W}}{m_W^2}= \frac{g_\text{NP}^2}{\Lambda^2}  \, ,
\eear
upto a factor $g$ from the conventional definition of $\mathcal{O}_W$. The value of $\Lambda$ corresponding to a value of $\bar{c}_W$ can be read off the upper x-axis in Fig.~\ref{fig:fitsummary} assuming $g^2_\text{NP}=1$, where we see that the marginalized range for $\bar{c}_W$ corresponds to $\Lambda \sim 400-800$ GeV. However $g_\text{NP}$ may vary to be less than 1 in weakly-coupled scenarios, in which case the new physics scale is lowered, or up to $4\pi$ for strongly-coupled new physics, which raises $\Lambda$. In general we have
\bear
\Lambda_{\bar c_W} \simeq \left(\frac{g_\text{NP}}{4 \pi}\right) \,  10 \textrm{ TeV.}
\label{lambda_cw}
\eear
The question can be asked whether the effective Standard Model approach is justified.

\begin{figure}[h!]
\centering
\includegraphics[scale=0.21]{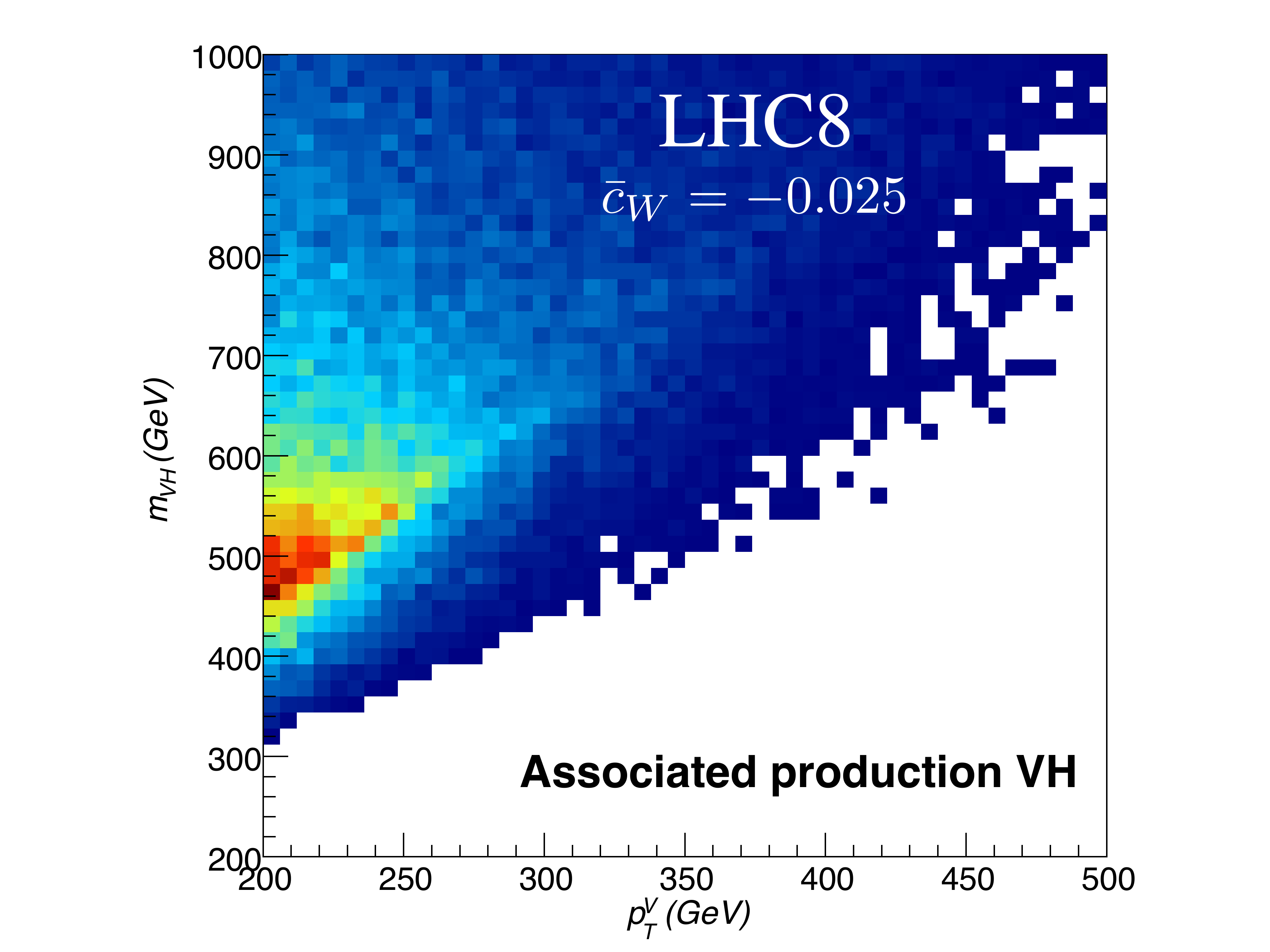}
\includegraphics[scale=0.21]{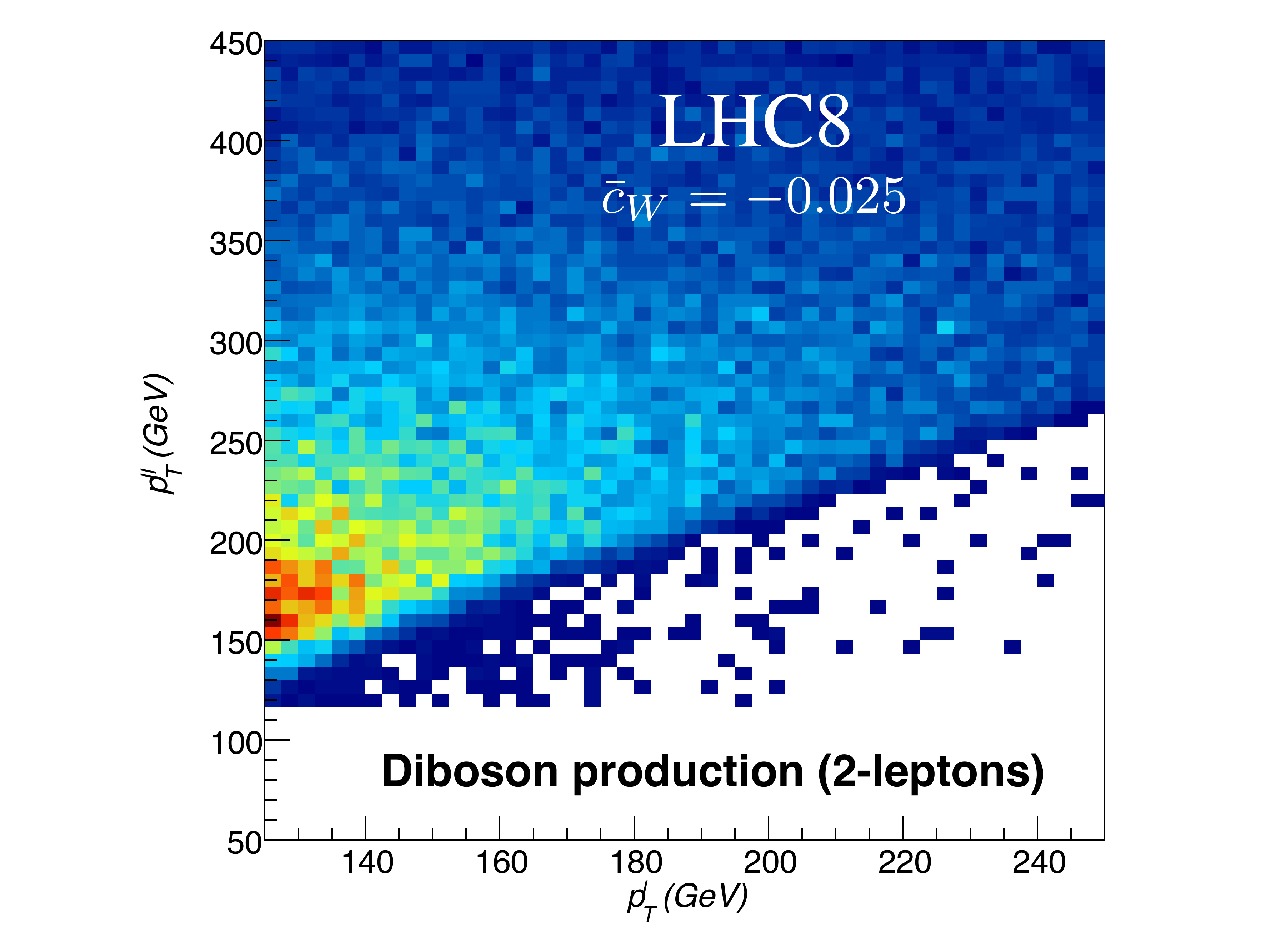}
\caption{\it (Left) The kinematic distribution in the vector boson $p_T^V$ vs $m_{VH}$ plane
for associated Higgs production at the LHC that would by induced by $\bar{c}_{W} = -0.025$.
(Right) The kinematic distribution in the leading lepton $p_T$ vs $p_T^{\ell\ell}$ plane
for diboson production at the LHC that would by induced by $\bar{c}_{W} = -0.025$.}
\label{fig:kinematics}
\end{figure}

In this Appendix we address this question by considering the region where the most sensitivity is obtained, i.e., the last bin.
First of all, it is important to note that the last bin is an overflow bin, containing all the events with $p_T$ above
a specified cut. For example, in the TGC analysis shown in
Fig.~\ref{fig:exampledistribution} the last bin corresponds to $p_T> 135$ GeV. 

For a given value of $\Lambda$, one expects the effective theory to break down at parton energies
$\sqrt{\hat s}\simeq \Lambda$, namely $m_{VV}$ and $m_{VH}$ in the diboson and VH production
respectively. To illustrate this point, in Fig.~\ref{fig:kinematics} we show the kinematic distribution that would be induced
by $\bar{c}_{W} = -0.025$ (our most conservative limit in $\bar c_W$)  in the plane defined by the transverse momentum
of the vector boson, $p_T^{V}$, and the invariant mass, $m_{VH}$,
for associated Higgs production at the LHC in the 2-lepton channel.
This plot corresponds to the last bin of the distribution, which has a cut $p_T^V>$ 200 GeV.
We see that in this bin typically $p_T^{V} \lesssim 250$~GeV, i.e., there is not a large spread of events
at large values of the distribution, and $\sqrt{\hat s}=m_{VH} \lesssim 550$~GeV. 

One can perform a similar analysis in the di-boson production case. 
For comparison, we show in the right panel of Fig.~\ref{fig:kinematics} the $p_T$ distribution of the 
leading lepton in the $p p \to W^+W^- \to 2 \ell+ \slashed{E}_T$ production at LHC8 versus the 
transverse mass distribution of the two vector bosons, $p_T^{ll}$. For comparison with
Fig.~\ref{fig:exampledistribution}, we infer that the overflow bin of $p_T>$ 135 GeV extends to about 160~GeV,
and is correlated with $p_T{\ell\ell} <$ 250 GeV.

Thus, in both the associated production and TGC cases, for $g_{NP} = {\cal O}(1)$,
equation (\ref{lambda_cw}) reassures us that the most important regions of the kinematical
distributions are well within the ranges where one may expect the effective field theory to be a good
enough approximation for our purposes.

% \bibliography{mono}{}
 \providecommand{\href}[2]{#2}\begingroup\raggedright

\end{document}